\documentclass[
    aps,
    prd,
    reprint,
    superscriptaddress,
    nobibnotes,
    nofootinbib,
]{revtex4-2}

\usepackage[dvipsnames, usenames]{xcolor}
\definecolor{linkcolor}{rgb}{0.0,0.3,0.5} %

\usepackage[
colorlinks=true,
linkcolor=linkcolor,
citecolor=linkcolor,
filecolor=linkcolor,
urlcolor=linkcolor,
pdfusetitle,
]{hyperref}
\usepackage{booktabs}
\usepackage{multirow}
\usepackage{amssymb}
\usepackage{amsmath}
\usepackage{physics}
\usepackage{orcidlink}
\usepackage{graphicx}
\usepackage{xcolor}
\usepackage{siunitx}
\usepackage{appendix}

\DeclareMathOperator*{\argmin}{arg\,min}

\newcommand{\Ac}{\mathcal{A}_{\times}}
\newcommand{\Ap}{\mathcal{A}_{+}}
\newcommand{\B}{\mathcal{B}}
\newcommand{\cost}{\mathcal{C}}
\newcommand{\costOne}{\langle\mathfrak{c}_1\rangle}
\newcommand{\dL}{d_{\mathrm{L}}}
\newcommand{\ETT}{\mathrm{ET}\triangle}
\newcommand{\ETL}{\mathrm{ET}2\mathrm{L}}
\newcommand{\hyp}{\mathcal{H}}
\newcommand{\Mchirp}{\mathcal{M}_{\mathrm{c}}}
\newcommand{\Tsft}{T_{\mathrm{SFT}}}
\newcommand{\Nsft}{N_{\mathrm{SFT}}}

\newcommand{\milan}{\affiliation{Dipartimento di Fisica ``G. Occhialini'', Universit\'a degli Studi di Milano-Bicocca, Piazza della Scienza 3, 20126 Milano, Italy}}
\newcommand{\infn}{\affiliation{INFN, Sezione di Milano-Bicocca, Piazza della Scienza 3, 20126 Milano, Italy}}
\newcommand{\lvn}{\affiliation{Centre for Cosmology, Particle Physics and Phenomenology - CP3,
Universit\'e Catholique de Louvain, Louvain-La-Neuve, B-1348, Belgium}}

\begin{document}

\title{Scalable early warning for binary neutron-star mergers in third-generation gravitational-wave detectors}

\author{Martin Gerini$\,$\orcidlink{0009-0005-5530-7522}}
\email{martin.gerini@uclouvain.be}
\lvn \milan

\author{Rodrigo Tenorio$\,$\orcidlink{0000-0002-3582-2587}}
\milan \infn

\author{Davide Gerosa$\,$\orcidlink{0000-0002-0933-3579}}
\milan \infn

\date{\today}

\begin{abstract}
Third-generation gravitational-wave observatories will observe binary neutron-star coalescences for hours before merger. Over these extended time windows, Earth's motion and the long signal durations make standard search methods computationally demanding for low-latency detection. We propose an approach based on short-time Fourier transforms and convolutional neural networks that implicitly marginalizes over extrinsic parameters, effectively reducing the search to the chirp mass. The network learns the signals' nonlinear time-frequency morphology, approximating the optimal Bayesian detection statistic. Using a large-scale campaign of simulated signal injections, we compare the network's performance with a GPU-optimized matched-filtering baseline for both the $\ETT$ and $\ETL$ configurations of the Einstein Telescope and signal durations from approximately 25 minutes ($f_{\min} = 5$ Hz) to approximately 6 hours ($f_{\min} = 2$ Hz). The network achieves comparable detection sensitivity while reducing computational costs by over four orders of magnitude, bringing processing times from minutes to milliseconds. This efficiency paves the way for rapid early-warning alerts for multi-messenger astronomy in the third-generation detector era.
\end{abstract}

\maketitle

\section{Introduction}

The detection of compact-binary coalescences by the ground-based LIGO--Virgo--KAGRA interferometer network~\cite{LIGOScientific:2014pky,VIRGO:2014yos,KAGRA:2018plz} has firmly established gravitational waves (GWs) as a new channel for studying the Universe~\cite{LIGOScientific:2026wfs}.
The sensitive band of the current, second-generation detector network is limited to frequencies $\gtrsim \SI{10}{\hertz}$ because of seismic and Newtonian noise~\cite{saulson_1994}.
This limited low-frequency sensitivity shortens the duration of GW signals, potentially hindering several scientifically important goals, such as the confident detection of intermediate-mass black holes~\cite{Reali:2024hqf} and the early warning, multi-messenger detection, and follow-up of binary neutron-star (BNS) mergers such as GW170817~\cite{LIGOScientific:2017vwq}.

Third-generation (3G) observatories, such as the Einstein Telescope (ET)~\cite{ET:2025xjr} in Europe and Cosmic Explorer~\cite{Reitze:2019iox} in the United States, will extend observations into the lower-frequency audio band of the GW spectrum, down to approximately \qtyrange{2}{5}{\hertz}.
Consequently, signals that are currently observed as short-duration transients will be observable for several hours before merger. %
Over these longer timescales, the observed GW signal is modulated by the Earth's rotation and orbital motion, invalidating common assumptions in GW data analysis~\cite{Dhurandhar:1992mw,Jaranowski:1998qm,Chen:2024kdc}.
Furthermore, long-duration signals require finer sampling of the parameter space to be properly characterized, increasing the computational cost of the analysis.
In practice, this can prevent low-latency searches from achieving the efficiency required for early-warning applications~\cite{Hu:2024mvn,Baker:2025taj}.

In this work, we present a pilot implementation based on deep learning, designed to address the scaling challenges associated with long-duration GW signals.
We process the signals using Short Fourier Transforms (SFTs) and train a convolutional neural network (CNN) to approximate the marginalized Bayes factor while reducing the dimensionality of the search parameter space to one.
This approach lowers the computational cost by a factor of $\sim 10^4$ while achieving sensitivity comparable to that of an equivalent matched-filtering search, thus enabling fast and scalable early-warning analyses.

\section{Matched-filtering as a classification problem}
\label{sec:mf_as_classification}

The standard approach to search for GW signals is to compare the observed data $d$ with a parameterized family of waveform models $h(t;\theta)$ and maximize a detection statistic%
~\cite{Allen:2002bp}.
In Bayesian hypothesis testing~\cite{Bretthorst_1988,Searle:2008jv,Prix:2009tq,Prix:2011qv}, this detection statistic corresponds to the Bayes factor between the signal and noise hypotheses,
\begin{equation}
    \B(d)
    = \frac{p(d | \hyp_{\mathrm{S}})}{p(d | \hyp_{\mathrm{N}})}
    = \int \mathrm{d}\theta \,
    \frac{p(d | \theta, \hyp_{\mathrm{S}})}{p(d | \hyp_{\mathrm{N}})}
    p(\theta | \hyp_{\mathrm{S}}) \,.
    \label{eq:BF}
\end{equation}
This integral is generally intractable.
In practice, the parameter space is divided into two subsets, $\theta = \{\lambda, \bar{\lambda}\}$, such that the marginalization over $\bar{\lambda}$ is performed analytically, while different values of $\lambda$ are evaluated numerically (see, e.g.,~\cite{Jaranowski:1998qm,Prix:2009tq,Tenorio:2024jgc}):
\begin{equation}
    \B(\lambda; d)
    = \int \mathrm{d}\bar{\lambda} \,
    \frac{p(d | \lambda, \bar{\lambda}, \hyp_{\mathrm{S}})}
    {p(d | \hyp_{\mathrm{N}})}
    p(\bar{\lambda} | \hyp_{\mathrm{S}}) \,.
    \label{eq:targeted_BF}
\end{equation}
The final detection statistic is then typically approximated by maximizing over $\lambda$~\cite{Usman:2015kfa,Tenorio:2021wmz}:
\begin{equation}
    \B(d) \sim \max_{\lambda} \B(\lambda; d) \,.
\end{equation}

The key insight is that the optimal detection statistic $\mathcal{B}$ for distinguishing between two complementary hypotheses is related to the optimal binary cross-entropy classifier~\cite{Cranmer:2015bka,Cranmer:2015bka}.
In this work, we recast Eq.~\eqref{eq:targeted_BF} in a simulation-based-inference framework, with the goal of replacing intractable integrals with a numerical minimization that is performed only \emph{once} for a given noise model and signal family.
in closed-form functions and reduce the computing cost of hard-to-evaluate detection statistics.

The construction proceeds as follows.
Let $r_{\phi}(d)$ be a scalar function of the data, parameterized by a set of parameters $\phi$, and define \mbox{$p_{\phi}(d) = \sigma(\ln r_{\phi}(d))$}. %
Here, \mbox{$\sigma(x) = [1 + e^{-x}]^{-1}$} is the sigmoid function.
Note that \mbox{$p(d | \hyp_{\mathrm{S}}) = \sigma(\ln \B(d))$} and, because the signal and noise hypotheses span the complete hypothesis space, \mbox{$p(d | \hyp_{\mathrm{N}}) = 1 - p(d | \hyp_{\mathrm{S}})$}.
Similarly, for an appropriate $\phi^{*}$,
\mbox{$p(d | \hyp_{\mathrm{S}}) \approx p_{\phi^{*}}(d)$} and
\mbox{$p(d | \hyp_{\mathrm{N}}) \approx 1 - p_{\phi^{*}}(d)$},
such that $r_{\phi^{*}}(d)$ is, in practice, equivalent to $\B(d)$.
It is straightforward to show that $\phi^{*}$ corresponds to the minimizer of the binary cross-entropy: 
\begin{align}
\phi^{*}
= \argmin_{\phi} \big\{
&-\langle \ln p_{\phi}(d) \rangle_{d \sim p(d | \hyp_{\mathcal{S}})}
\notag \\
&
-\langle \ln[1 - p_{\phi}(d)] \rangle_{d \sim p(d | \hyp_{\mathcal{N}})}
\big\} \,.
\label{eq:BCE}
\end{align}
Therefore, the Bayes factor in Eq.~\eqref{eq:BF} can be computed by training a classifier $r_{\phi}$ that minimizes the binary cross-entropy loss in Eq.~\eqref{eq:BCE}.
In practice, this converts a high-dimensional and intractable integral, which standard searches must evaluate numerically for all waveforms $\theta$~\cite{Capano:2013raa,Capano:2016dsf,Usman:2015kfa,Islam:2022afg,Wadekar:2024zdq}, into a minimization problem that can be solved offline for a given noise model $p(d | \hyp_{\mathrm{N}})$ and signal distribution $p(d | \hyp_{\mathrm{S}})$.

\section{Application to early-warning of BNS mergers in 3G detectors}
\label{earlywarning}

The short duration of GW signals in current detectors allows search pipelines (e.g.,~\cite{Usman:2015kfa,2021SoftX..1400680C}) to neglect the motion of the Earth relative to the Solar System barycenter (SSB).
As a result, the detector antenna-pattern functions are treated as constant, and standard template banks are typically restricted to the component masses and aligned-spin components of the two compact objects.
The broader sensitivity band of 3G detectors, however, is expected to invalidate this assumption~\cite{Chen:2024kdc}, as GW signals will be affected by time-dependent amplitude and phase modulations caused by the motion of the detectors, similarly to continuous GW signals in current detectors~\cite{Jaranowski:1998qm}. 
This will require expanding the search space to include additional parameters, such as the sky location %
and polarization angles. %
The longer signal duration will also increase the number of templates required to adequately cover the parameter space~\cite{Owen:1995tm}.
Overall, the computational cost is expected to increase to the point of hindering low-latency operations, such as early-warning searches,  affecting the multi-messenger science case of 3G detectors~\cite{Hu:2024mvn,Baker:2025taj}.

Leveraging Eq.~\eqref{eq:targeted_BF}, we construct a detection statistic for BNS early-warning searches as a function of the chirp mass, $\lambda = \{\Mchirp\}$, by training a classifier that effectively marginalizes over the remaining variables.
This reduces the search to a one-dimensional parameter space, requiring significantly fewer templates to achieve adequate coverage.

This strategy distinguishes our framework from previous applications of deep learning to GW searches, in which matched-filtering has been replaced with end-to-end classifiers~\cite{Gabbard:2017lja,Nagarajan:2025hws}.
Although these approaches have shown promise for short-duration transients in current detectors, and for BNS early warning on frames of a few
minutes~\cite{Baltus:2022pep}, replacing physical modeling with a neural network scales poorly to long-duration signals because the network must implicitly learn a high-dimensional parameter space, potentially introducing severe learning biases~\cite{Nagarajan:2025hws,Joshi:2023hpx}.
Other successful approaches in the field restrict machine learning to the post-processing of intermediate data products, which are treated as carefully crafted summary statistics~\cite{Yamamoto:2020pus,Modafferi:2023nzt}.
By construction, however, this strategy limits the extent to which search pipelines can be simplified and accelerated.

Our proposal targets the core of the problem---namely, the computation of the optimal detection statistic---by encapsulating computationally expensive operations, such as marginalization over the parameter space, in a simple variational problem that can be solved numerically.

\subsection{Overview}
\label{sec:data_gen}

We simulate the early inspiral of a BNS signal using the \texttt{IMRPhenomT}~\cite{Estelles:2020osj} approximant as implemented in \texttt{iphenot}~\cite{Tenorio:2025gci}.
We consider $f_{\text{min}} = \{2,3,4,5\} \, \mathrm{Hz}$, consistent with the expected low-frequency sensitivity of ET~\cite{ET:2019dnz}, and set the maximum frequency to $f_{\mathrm{max}} = 10 \, \mathrm{Hz}$~\cite{Tsutsui:2020bem}.
The two GW strain polarizations, $h_{+,\times}$, are generated in the SSB frame and projected onto the interferometric detectors.
This projection explicitly includes the amplitude and phase modulations induced by the Earth's motion and the anisotropic antenna-pattern functions~\cite{Jaranowski:1998qm} (see Appendix~\ref{app:signal_model}).
The sampling interval is $\Delta t = 0.04 \, \mathrm{s}$ for $f_{\mathrm{min}} = 2 \, \mathrm{Hz}$ and $\Delta t = 0.025 \, \mathrm{s}$ for $f_{\mathrm{min}} \in \{3,4,5\} \, \mathrm{Hz}$, balancing the available computational resources against the different waveform durations.

The resulting strain $h(t)$ is processed using SFTs~\cite{Tenorio:2025gci} by dividing it into $\Nsft$ disjoint time segments of duration $\Tsft = 31 \, \text{s}$, producing the complex SFT $\tilde{h}_{\alpha}(f_k)$ for each segment:
\begin{equation}
\tilde{h}_{\alpha}(f_k)
= \Delta t \sum_{j=0}^{n_{\mathrm{SFT}}-1}
h(t_{\alpha} + j \Delta t)
e^{-i 2\pi j k / n_{\mathrm{SFT}}} \,,
\label{eq:h_SFT}
\end{equation}
where $t_\alpha$ denotes the initial time of segment $\alpha$, \mbox{$n_{\mathrm{SFT}} = \operatorname{floor}(\Tsft / \Delta t)$}, and $f_k = k / \Tsft$.
This time-frequency representation of the strain is added to an independent realization of stationary Gaussian noise colored by the power spectral density $S_{\mathrm{n}}$ used in the Cost and Benefit Analysis (CoBA) Science study~\cite{Danilishin:2023ETcurves, Branchesi:2023mws}, producing the final data stream
\begin{equation}
\tilde{d}_{\alpha}(f_k)
= \tilde{h}_{\alpha}(f_k) + \tilde{n}_{\alpha}(f_k) \,.
\end{equation}
The real and imaginary components of the noise are sampled from zero-mean Gaussian distributions with identical standard deviations~\cite{Tenorio:2024jgc}: 
\begin{equation}
\operatorname{Re}\tilde{n}_{\alpha}(f_k),
\operatorname{Im}\tilde{n}_{\alpha}(f_k)
\sim
\mathcal{N}\left(
0, \frac{\sqrt{\Tsft S_{\mathrm{n}}(f_k)}}{2}
\right) \,.
\end{equation}
We refer to $\tilde{d}_{\alpha}(f_k)$ as a {complex spectrogram}, in the sense that $|\tilde{d}_{\alpha}(f_k)|^2$ corresponds to the conventional spectrogram of the data.

\begin{figure}
    \centering
    \includegraphics[width=1\linewidth]{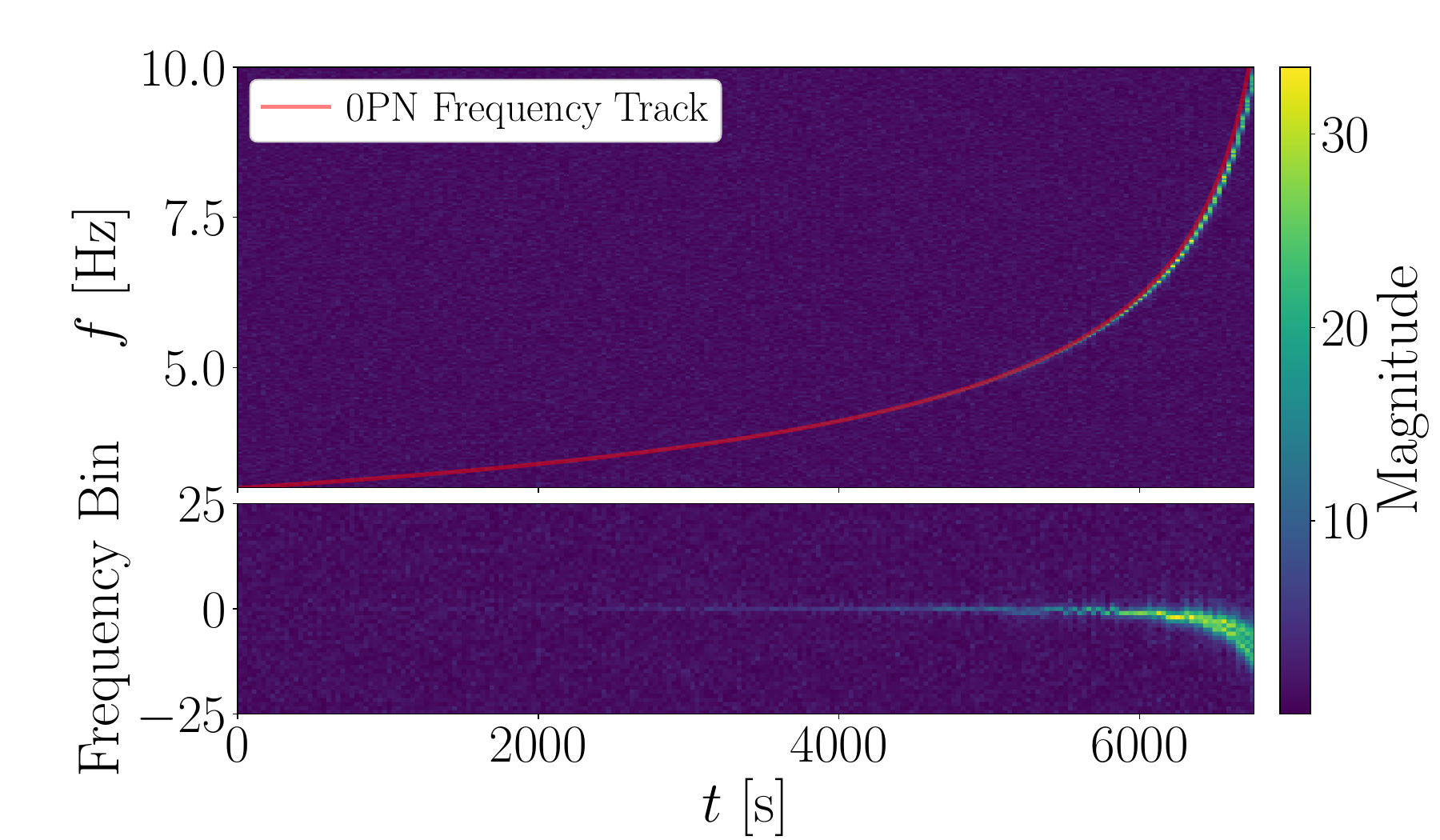}
    \caption{Whitened GW signal from a BNS system with $m_1 = m_2 = 3 \, M_\odot$ at $d_L = 100 \, \mathrm{Mpc}$. 
    \emph{Upper}: Spectrogram of a simulated GW signal (yellow track) overlaid with the leading-order frequency track (red line). \emph{Lower}: Cropped SFT spectrogram centered on the 0PN track.
    }
    \label{fig:bns_track}
\end{figure}

During the early inspiral, the GW signal from a BNS can be described using a quasi-monochromatic model.
This suggests that an effective detection statistic can be constructed by focusing on a narrow region of the complex spectrogram, as argued in Ref.~\cite{Tenorio:2025gci} and illustrated in Fig.~\ref{fig:bns_track}.
Therefore, rather than using the complete complex spectrogram, we use narrow bands of $\Delta k = 51$ frequency bins centered on a theoretical frequency track as input to $\ln r_\phi$.
We model this track using an analytical Newtonian approximation determined by the chirp mass~\cite{Maggiore:2007ulw}:
\begin{equation}
\begin{split}
    &f(t;\Mchirp) = \\&f_\text{max} 
    \left[1 + \frac{256}{5}(\pi f_\text{max})^{8/3} \left( \frac{G \Mchirp}{c^3}\right)^{5/3} (t_\text{max} - t)\right]^{-3/8} \,,
\end{split}
\end{equation}
where $0 \leq t \leq t_\text{max}$.
We evaluate the frequency backward in time from a fiducial reference time $t_\text{max}$, corresponding to the moment at which the signal reaches a predetermined maximum frequency $f_\text{max}$ within the data segment.

\subsection{Network architecture and data pre-processing}

We model $\ln r_\phi$ using a CNN.
The architecture comprises four convolutional blocks, whose output is flattened and passed through a multilayer perceptron.
The final layer applies a linear transformation without an activation function to improve numerical stability.
Further details are provided in Appendix~\ref{appendix:CNNS}.

For each time segment $\alpha$, we identify the central frequency bin,
\mbox{$k_\alpha = \operatorname{int}[f(t_\alpha;\Mchirp)\Tsft]$}, %
and extract a localized bandwidth of $2P+1$ bins:
\begin{equation}
    d(t_{\alpha}; \Mchirp) = \left[ \tilde{d}_{\alpha}[k_\alpha - P], \dots, \tilde{d}_{\alpha}[k_\alpha + P] \right] \in \mathbb{C}^{2P+1} \,.
\end{equation}
Because $\ln r_{\phi}$ takes real-valued inputs, we separate the real and imaginary parts of $d(t_{\alpha};\Mchirp)$ into distinct channels for each detector:
\begin{align}
    \mathbf{D}_{\mathrm{det}}(\Mchirp)
    &= \left[\mathrm{Re}\, d(t_{\alpha}; \Mchirp), \mathrm{Im}\, d(t_{\alpha}; \Mchirp) \right], \\
    \mathbf{D}_{\mathrm{det}}(\Mchirp) &\in \mathbb{R}^{2 \times (2P+1) \times N_{\text{SFT}}} \,.
\end{align}
For multiple detectors, we stack the corresponding data arrays:
\begin{align}
    \mathbf{D}(\Mchirp) &= [\mathbf{D}_{\mathrm{det}}(\Mchirp) , \mathrm{det}=1, \dots, N_{\mathrm{det}}],\\
    \mathbf{D}(\Mchirp)&\in \mathbb{R}^{2N_{\mathrm{det}} \times (2P+1) \times N_{\text{SFT}}}  \,.
\end{align}
Finally, the data array is whitened using the power spectral density of the noise-generating process.

\subsection{Detection pipeline}

The search procedure uses a template bank spanning chirp mass,
$\mathcal{T}_{\text{CNN}} = \{\Mchirp^{(l)}, l=0, \dots, \mathcal{N}-1\}$.
Templates are generated by uniformly sampling the target astrophysical range $\Mchirp \in [0.87, 2.61] \, \mathrm{M_\odot}$, which corresponds to BNS systems with component masses $m_1, m_2 \in [1.0, 3.0] \, \mathrm{M_\odot}$.

For each template $\Mchirp^{(l)}$, the corresponding data array $\mathbf{D}(\Mchirp^{(l)})$ is fed into the CNN, whose output is interpreted as a targeted detection statistic, $\ln r^{(l)}_{\phi} = \ln r_{\phi}[\mathbf{D}(\Mchirp^{(l)})]$, as described in Sec.~\ref{sec:mf_as_classification}.
The final detection statistic, corresponding to Eq.~\eqref{eq:BF}, is obtained by taking the maximum value across the template bank:
\begin{equation}
s_{\text{CNN}} = \max_{l} \ln r_{\phi}^{(l)} \,.
\end{equation}

\subsection{Training}
\label{sec:training}

The noise distribution $p(d |\hyp_{\mathrm{N}})$ is modeled independently for each detector as stationary Gaussian noise, consistent with the ET sensitivity curves used for CoBA Science Study~\cite{Danilishin:2023ETcurves}.

For the signal distribution $p(d | \hyp_{\mathrm{S}})$, the component masses and aligned spins are sampled uniformly, with \mbox{$m_1, m_2 \sim \mathcal{U}(1.0, 3.0)\,M_\odot$} and \mbox{$\chi_{1L}, \chi_{2L} \sim \mathcal{U}(-0.05, 0.05)$}~\cite{LIGOScientific:2017vwq,Burgay:2003jj}.
The source's sky location $(\alpha, \delta)$, polarization angle $\psi$, and inclination $\iota$ are sampled isotropically, i.e., $\alpha \sim \mathcal{U}(0, 2\pi)$, $\sin\delta \sim \mathcal{U}(-1, 1)$, $\psi \sim \mathcal{U}(-\pi/2, \pi/2)$, and $\cos\iota \sim \mathcal{U}(-1, 1)$.
The luminosity distance is sampled uniformly as $d_L \sim \mathcal{U}(d_{\text{min}}, d_{\text{max}})$, corresponding to a redshift range of $z \in [0.01, 0.15]$ under the \textsc{Planck15} cosmology~\cite{Planck:2015fie}.
Finally, the start time $t_0$ is drawn uniformly over a full Julian year, and the initial phase is sampled as $\phi_0 \sim \mathcal{U}(0, 2\pi)$.

We use the loss function of Eq.~\eqref{eq:BCE} 
to train the network with a training set generated on-the-fly:
at each training step, we sample a batch of  \mbox{$n_{\mathrm{batch}} = 11$} parameters 
$\{\theta^{(b)}, b=1, \dots, n_{\mathrm{batch}}\}$ from the prior distribution 
$p(\theta|\hyp_{\mathrm{S}})$.
For each $\theta$, we generate a pair of training samples. 
For $\hyp_{\mathrm{S}}$, the signal $h(t; \theta)$ is injected into a noise realization and
the corresponding track $f[t;\Mchirp(\theta)]$ is extracted with a bandwidth of $\Delta k = 51$.
For $\hyp_{\mathrm{N}}$, a noise sample is generated and a narrow bandwidth is extracted using the
same $f[t;\Mchirp(\theta)]$ track. This ensures that the background noise of noise samples is consistent
with that of signal samples.

\subsection{Matched-filtering search}
\label{sec:mf_baseline}

To compare the performance of our CNN, we set up a matched-filtering search based on the SFT
framework discussed in~\cite{Tenorio:2025gci} .
The matched-filtering search operates across a uniform 8-dimensional parameter space defined by 
$\theta = \{ m_1, m_2, \chi_{1L}, \chi_{2L}, \alpha, \delta, \psi, \cos\iota \}$. 
The final detection statistic $s_{\mathrm{SFT}}$ is obtained as the highest detection 
matched filtering evaluation across the template bank.
See Appendix~\ref{sft} for details.

\subsection{Evaluation setup}
We compare the detection efficiency of the proposed CNN with that of the baseline SFT pipeline through an injection campaign using simulated signals, following an approach similar to that of Refs.~\cite{Oliver:2018gbu,Tenorio:2024jgc}.
Specifically, we evaluate the detectability of a population of BNS systems at different luminosity distances $\dL$ and report $\dL^{90\%}$, i.e., the luminosity distance at which 90\% of the systems are successfully detected.

The fraction of detected systems estimates the detection probability $p_{\mathrm{det}}$, which we define as the probability that a signal's detection statistic exceeds a predefined threshold $\tau_{\text{fa}}$:
\begin{equation}
    p_{\text{det}} = p(s > \tau_{\text{fa}}|\hyp_{\mathrm{S}})\,.
\end{equation}
This threshold is set by fixing the false-alarm probability (FAP), which is the probability that a noise sample is classified as a signal:
\begin{equation}
    \text{FAP} = p(s > \tau_{\text{fa}}| \hyp_{\mathrm{N}}) \,.
\end{equation}
Because the signal duration $T_{\mathrm{sig}}$ depends strongly on the low-frequency cutoff (see Appendix~\ref{appendix:signal_lenght} for details), we convert the fixed FAP into a false-alarm rate (FAR), which measures the expected number of false alarms per unit time~\cite{Morras:2022ysx}:
\begin{equation}
    \mathrm{FAR} = \frac{-\ln(1-\text{FAP})}{T_{\mathrm{sig}}}\,.
\end{equation}

The injection campaign span a parameter grid defined by:
\begin{enumerate}
\item Luminosity distance $\dL$. To evaluate the sensitivity depth, we vary the luminosity distance of the injected sources.
We use 13 distance values corresponding to redshifts $z \in [0.1, 0.5]$, computed using the \textsc{Planck15} cosmology.\cite{Planck:2015fie}. %
    \item Bank Size $\mathcal{N}$. To evaluate the impact of sampling density, 
    we tested various template bank sizes. 
    The ranges differ significantly between the two methods due to the 
    intrinsic dimensionality of their respective search spaces:
    \begin{align}
        \mathcal{N}_{\text{SFT}} &\in \{10^3, 10^4, 10^5, 10^6, 10^7\}\,, \\
        \mathcal{N}_{\text{CNN}} &\in \{10^2, 10^3, 10^4\}\,.
    \end{align}
    \item Detector configuration. We consider the two currently discussed Einstein Telescope configurations~\cite{Branchesi:2023mws}: 
    three co-located detectors with an opening angle of \mbox{$\zeta = 60^{\circ}$}, which we place
    in the current Advanced Virgo site~\cite{VIRGO:2014yos} ($\ETT$), 
    and two \mbox{$\zeta = 90^{\circ}$} detectors with a relative orientation of $45^{\circ}$, separated by about 
    \SI{1000}{\kilo\meter} ($\ETL$), which we locate at the Advanced Virgo site and 
    the GEO600 site~\cite{Willke:2002bs}.
    Each layout is modeled with the ETLF CoBA sensitivity
    curve~\cite{Danilishin:2023ETcurves} corresponding to its arm length,
    $10\,\text{km}$ for $\ETT$ and $15\,\text{km}$ for $\ETL$.
    Within a layout, all interferometers share the same PSD, and the noise is
    assumed to be Gaussian and uncorrelated between interferometers.

\end{enumerate}
We estimate FAP and $p_{\mathrm{det}}$ by drawing $10^3$ independent noise statistics $s \sim p(s \mid \hyp_{\mathrm{N}})$ and $10^3$ independent signal statistics $s \sim p(s \mid d_{\mathrm{L}}, \mathcal{N}, \mathrm{detector}, \hyp_{\mathrm{S}})$ for each combination of $(d_{\mathrm{L}}, \mathcal{N}, \mathrm{detector})$.
We set the detection threshold $\tau_{\text{fa}}$ to the value corresponding to a false-alarm probability of $\text{FAP} = 1/N_{\text{iter}} = 0.1\%$, which yields acceptable FARs, as shown in Table~\ref{tab:far_values}.

\begin{table}
    \begin{ruledtabular}
    \begin{tabular}{cccc}
        $f_{\min} \, [\mathrm{Hz}]$  & $T_{\mathrm{sig}} \, [\mathrm{s}]$ & $\mathrm{FAR} \,[\mathrm{Hz}]$ & $\mathrm{FAR}^{-1}\,[\mathrm{month}]$ \\
        \colrule
        \rule{0pt}{3ex}2 & $20\,460$ & $4.89 \times 10^{-8}$ & $7.8$ \\
        3 & $6\,758$  & $1.48 \times 10^{-7}$ & $2.6$ \\
        4 & $2\,976$  & $3.36 \times 10^{-7}$ & $1.1$ \\
        5 & $1\,519$  & $6.59 \times 10^{-7}$ & $0.6$ \\
    \end{tabular}
    \end{ruledtabular}
    \caption{Signal durations $T_{\mathrm{sig}}$, corresponding FAR, 
    and the expected time intervals between FARs (i.e. $\mathrm{FAR}^{-1}$)
    for a false alarm probability $\text{FAP} = 0.1\%$.}
    \label{tab:far_values}
\end{table}

The sensitivity distance $\dL^{90\%}$ is obtained by fitting the empirical efficiency curves $p_{\text{det}}(d_L)$ at five selected distances near $\dL^{90\%}$ using a sigmoid function of the logarithmic distance:
\begin{equation}
    S(\dL;\ a,\ b) = \left[1 + \exp\left({\frac{\log_{10}\dL - b}{a}}\right)\right]^{-1}
\end{equation}
The free parameters $(a,b)$ are determined through a weighted least-squares fit, assuming binomial uncertainties in the estimated detection probabilities~\cite{Oliver:2018gbu,Tenorio:2024jgc}.

\section{Results}
\label{sec:results}

\subsection{Sensitivity}
\label{subsec:sensitivity}

\begin{figure}
    \centering
    \includegraphics[width=1\linewidth]{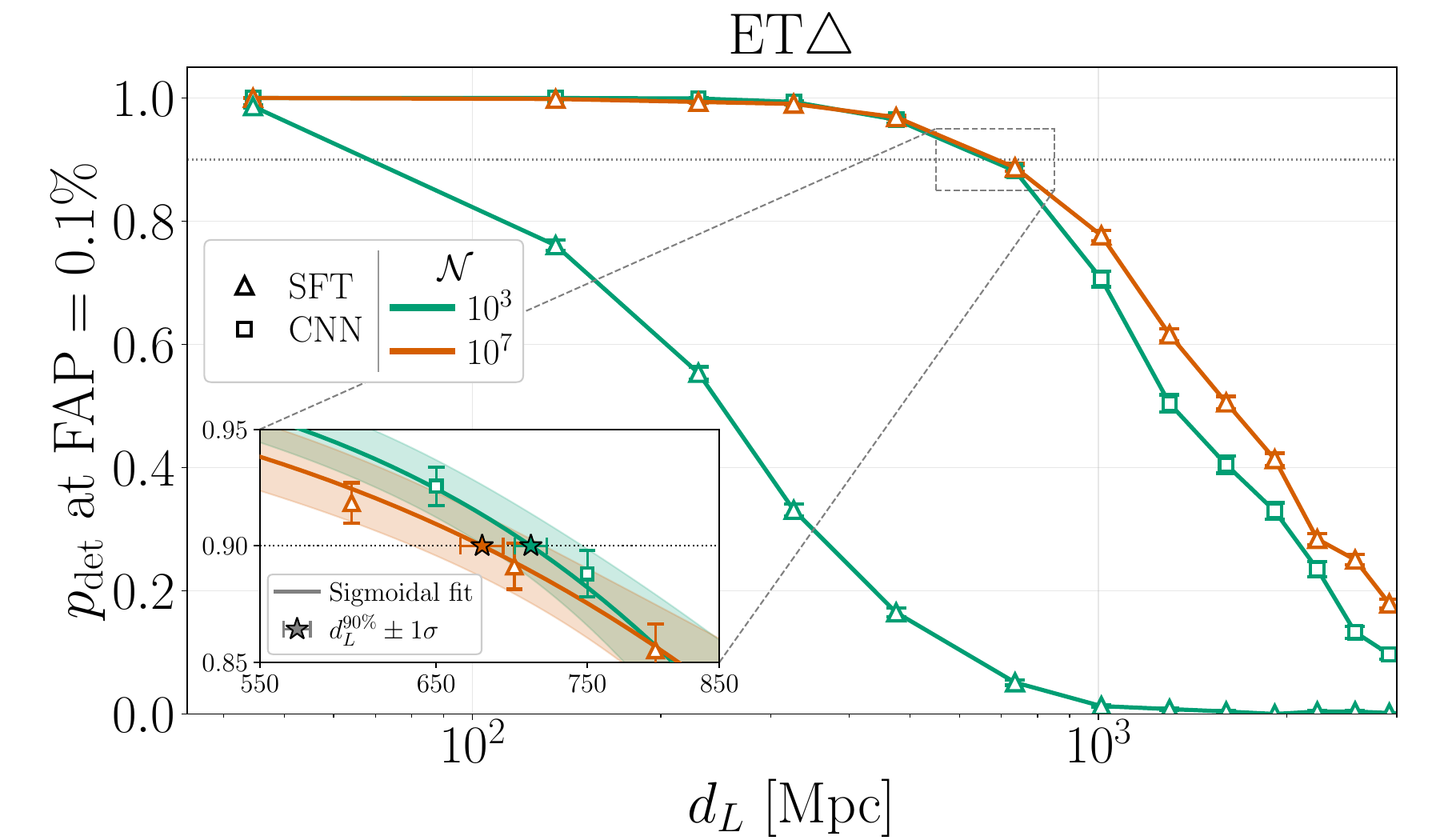}\\
    \vspace{0.4cm}
    \includegraphics[width=1\linewidth]{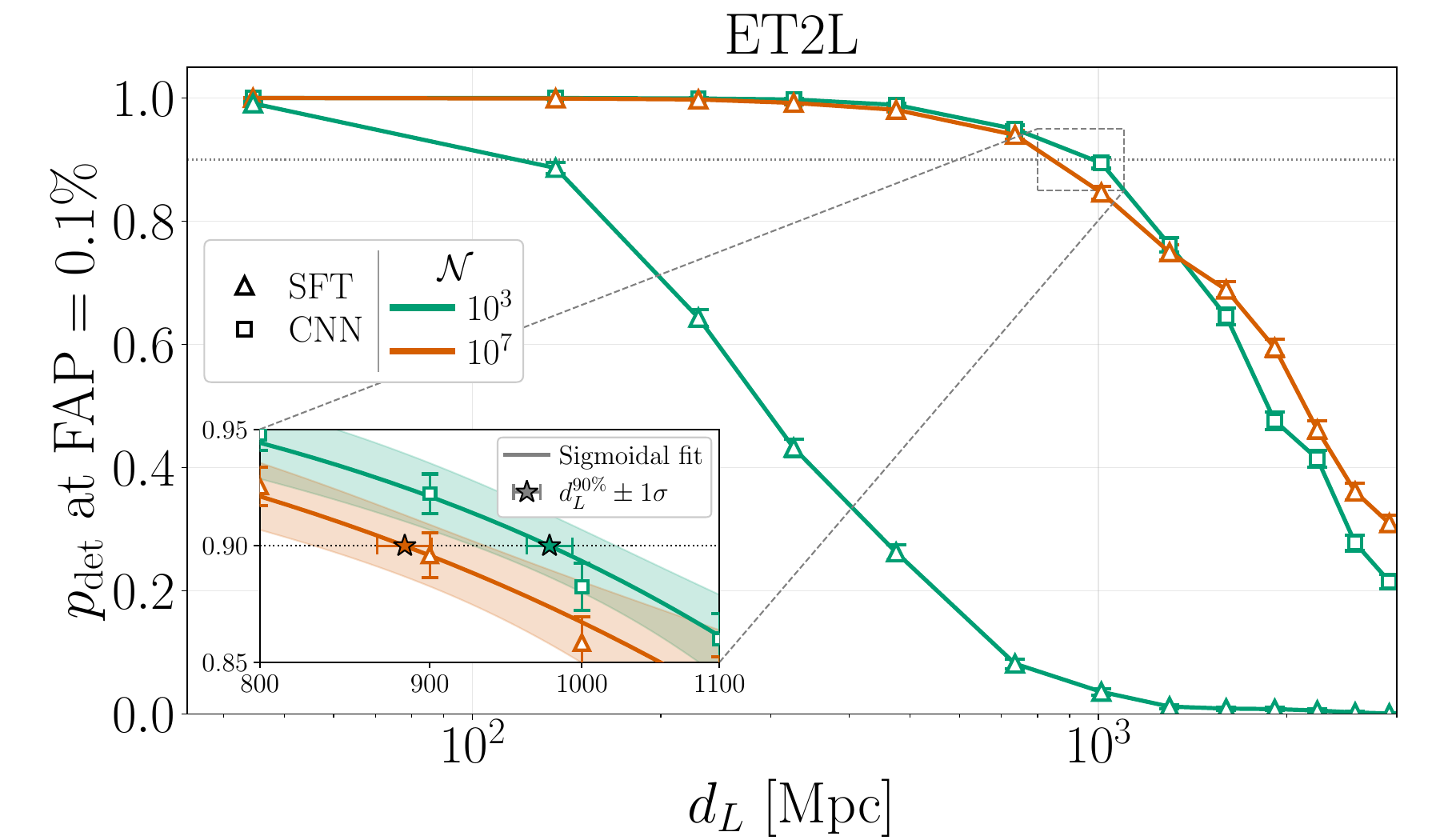}
    \caption{
    Detection probability $p_{\text{det}}$ as a function of the injected luminosity distance $d_L$, evaluated at a fixed false-alarm probability $\text{FAP}=0.1\%$. The plots illustrate the performance for the $\ETT$ (top) and $\ETL$ (bottom) configurations at a low-frequency cutoff of $f_{\text{min}} = 2$ Hz. 
    The proposed CNN pipeline operating at its saturation point 
    ($\mathcal{N} = 10^3$ templates, green squares) 
    is compared against the SFT matched filtering baseline with the same number of templates ($\mathcal{N} = 10^3$, green triangles) and with the densest template bank used in this study ($\mathcal{N} = 10^7$ templates, orange triangles). 
    The insets provide a magnified view around $\dL^{90\%}$. 
    Solid lines represent the sigmoidal fits used to extract the $\dL^{90\%}$ results 
    (marked by stars), with shaded region denoting the $3\sigma$ confidence band. 
    Error bars on discrete data points indicate $1\sigma$ binomial uncertainties.
    }
    \label{fig:sensitivity_example_2hz}
\end{figure}

\begin{figure}
    \centering
    \includegraphics[width=\columnwidth]{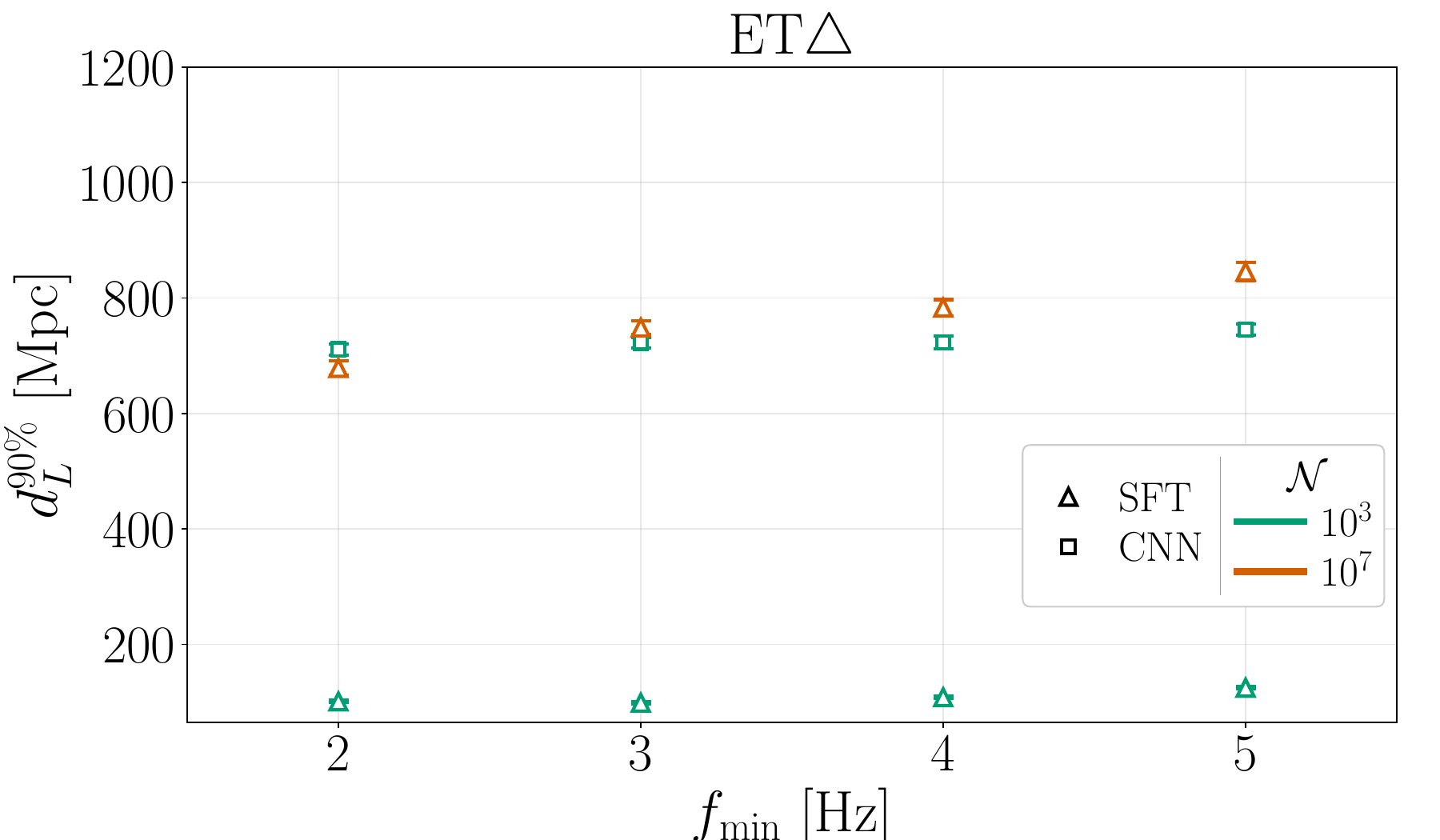}\\
    \vspace{0.4cm}
    \includegraphics[width=\columnwidth]{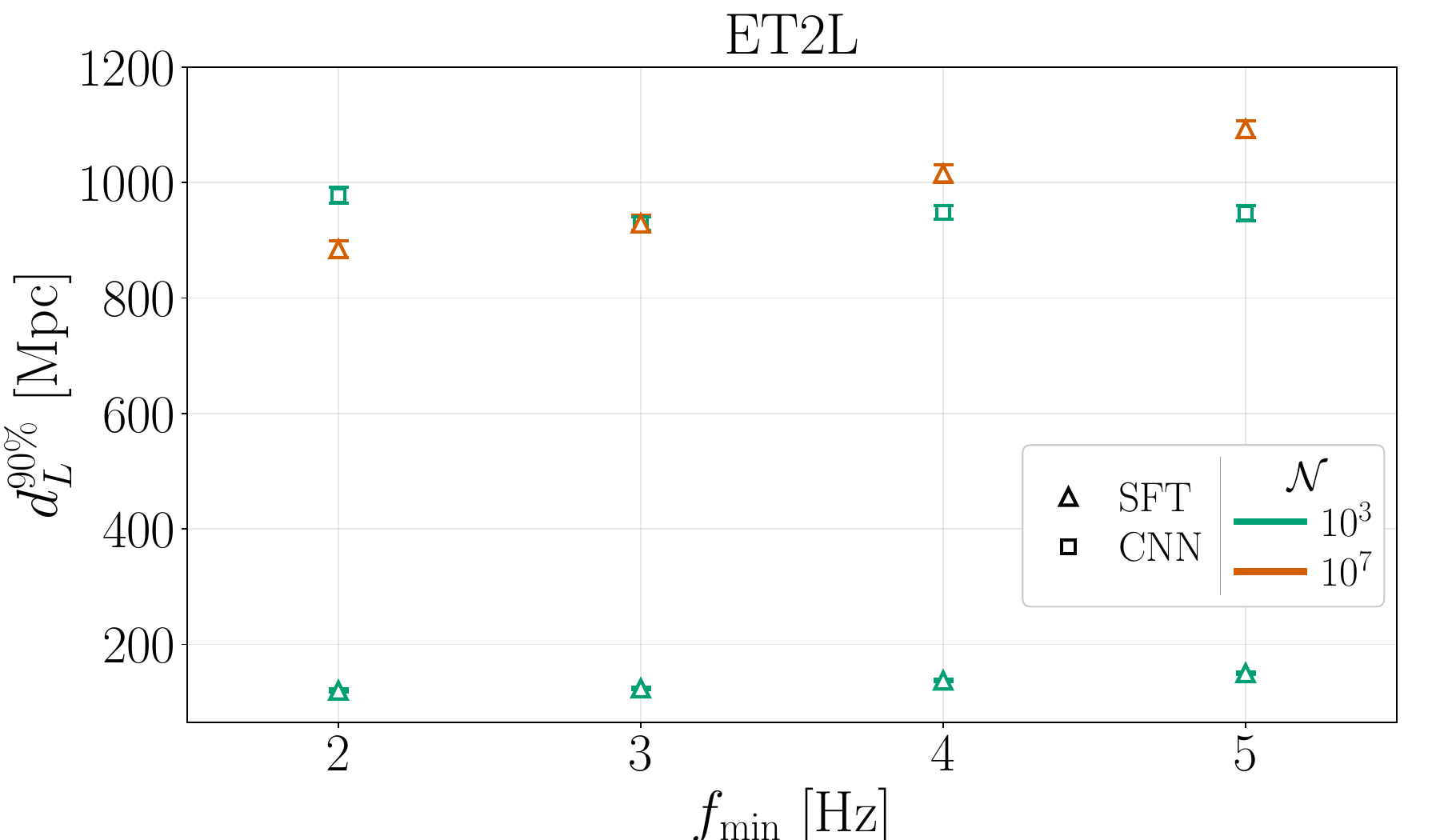}
    \caption{Detection sensitivity horizons $d_L^{90\%}$ evaluated at $\text{FAP} = 0.1\%$ for the $\ETT$ (top) and $\ETL$ (bottom) configurations across the 4 minimum frequency regimes considered in this study.
    The CNN pipeline (squares) employing $\mathcal{N} =10^3$ templates (green) is compared against the SFT matched filtering (triangles) operating with the same number of templates and with $\mathcal{N} = 10^7$ templates (orange). Error bars indicate $1\sigma$ binomial uncertainties.}
    \label{fig:d90vsfreq}
\end{figure}

An example sensitivity curve $(\dL, p_{\mathrm{det}})$ for both $\ETT$ and $\ETL$ with $f_\text{min} = 2$ Hz is shown in Fig.~\ref{fig:sensitivity_example_2hz}.
We find that the performance of the CNN approach saturates at $\mathcal{N}_{\text{CNN}} = 10^3$ templates, and outperforms matched-filtering operating with the same number of templates ($\mathcal{N}_{\text{SFT}} = 10^3$).
To achieve a comparable sensitivity distance, the matched-filtering approach requires approximately $\mathcal{N}_{\mathrm{SFT}} = 10^{7}$ templates, about four orders of magnitude more.
The corresponding estimates of $\dL^{90\%}$ at these two operating points are shown in Fig.~\ref{fig:d90vsfreq} for each detector configuration and minimum frequency.
For instance, in the $\ETL$ configuration at $2$ Hz, the CNN yields a sensitivity distance of $\dL^{90\%} = 978 \pm 14 \,\mathrm{Mpc}$, while the corresponding matched-filtering result is $\dL^{90\%} = 885 \pm 15 \,\mathrm{Mpc}$ with $\mathcal{N}_{\text{SFT}} = 10^7$ templates.
A similar trend is observed across all other frequency ranges and detector configurations, with the CNN approach achieving at least $86\%$ of the matched-filtering sensitivity distance.

\subsection{Latency and computing cost}

We model the latency of an early-warning pipeline, $\mathcal{C}$, as the cost of evaluating the detection statistic:
\begin{equation}
    \cost = \mathcal{N} \costOne \,,
\end{equation}
where $\mathcal{N}$ is the number of templates in the bank and $\costOne$ is the average cost of evaluating one template.
In practice, we evaluate multiple templates simultaneously using JAX's \texttt{vmap} function~\cite{jax2018github} and divide the resulting wall-clock time by the batch size.
We evaluated both the CNN and the matched-filtering approach on a dedicated computing node
with a single NVIDIA A100-SXM-64GB GPU. The reported costs are averaged across 10 realizations
of $N_{\text{sim}} = 1000$ independent template batches. These measurements were recorded after
an initial warm-up phase to remove overheads.

\begin{figure}
    \centering
    \includegraphics[width=1\linewidth]{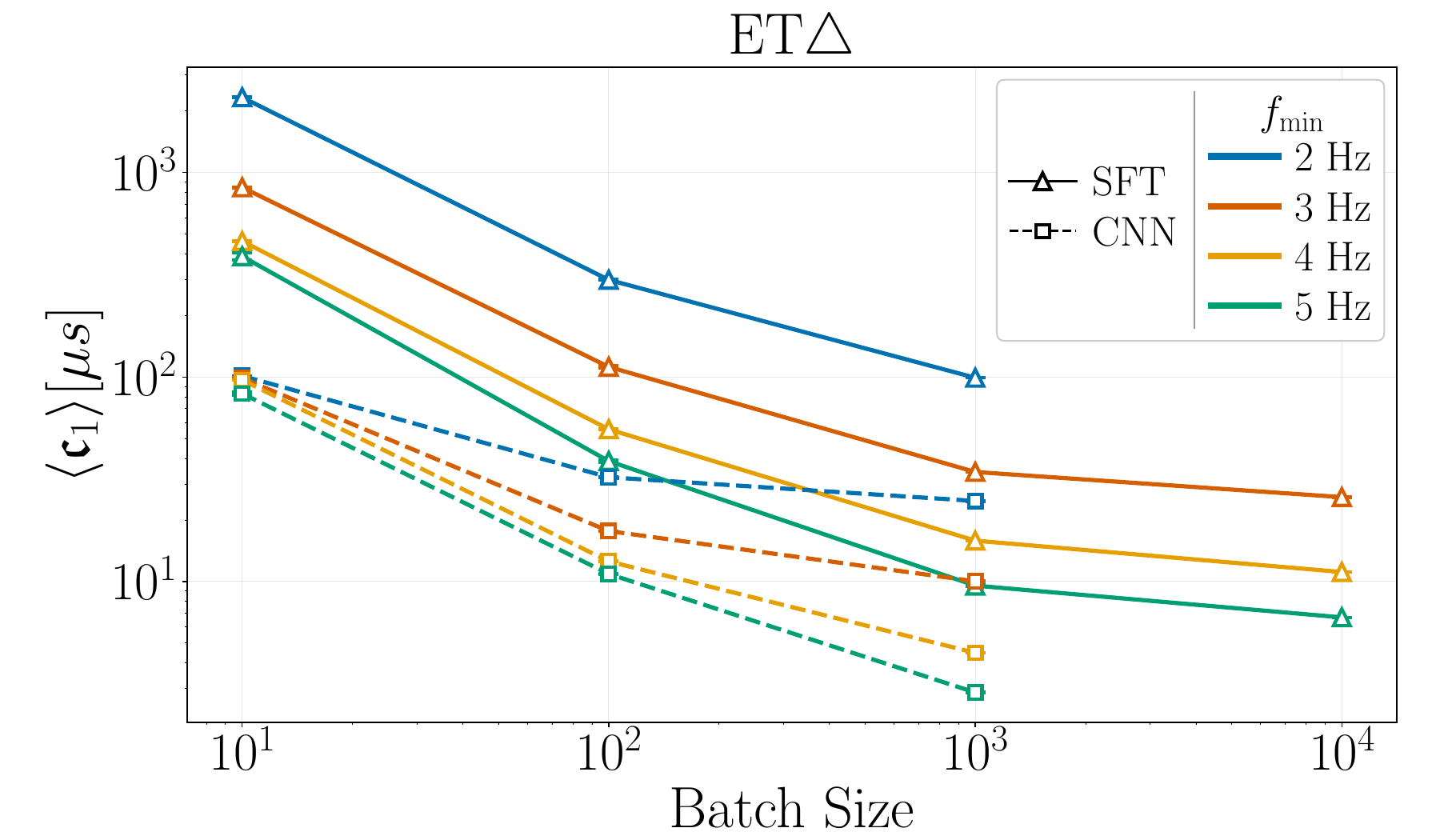}\\
    \vspace{0.4cm} %
    \includegraphics[width=1\linewidth]{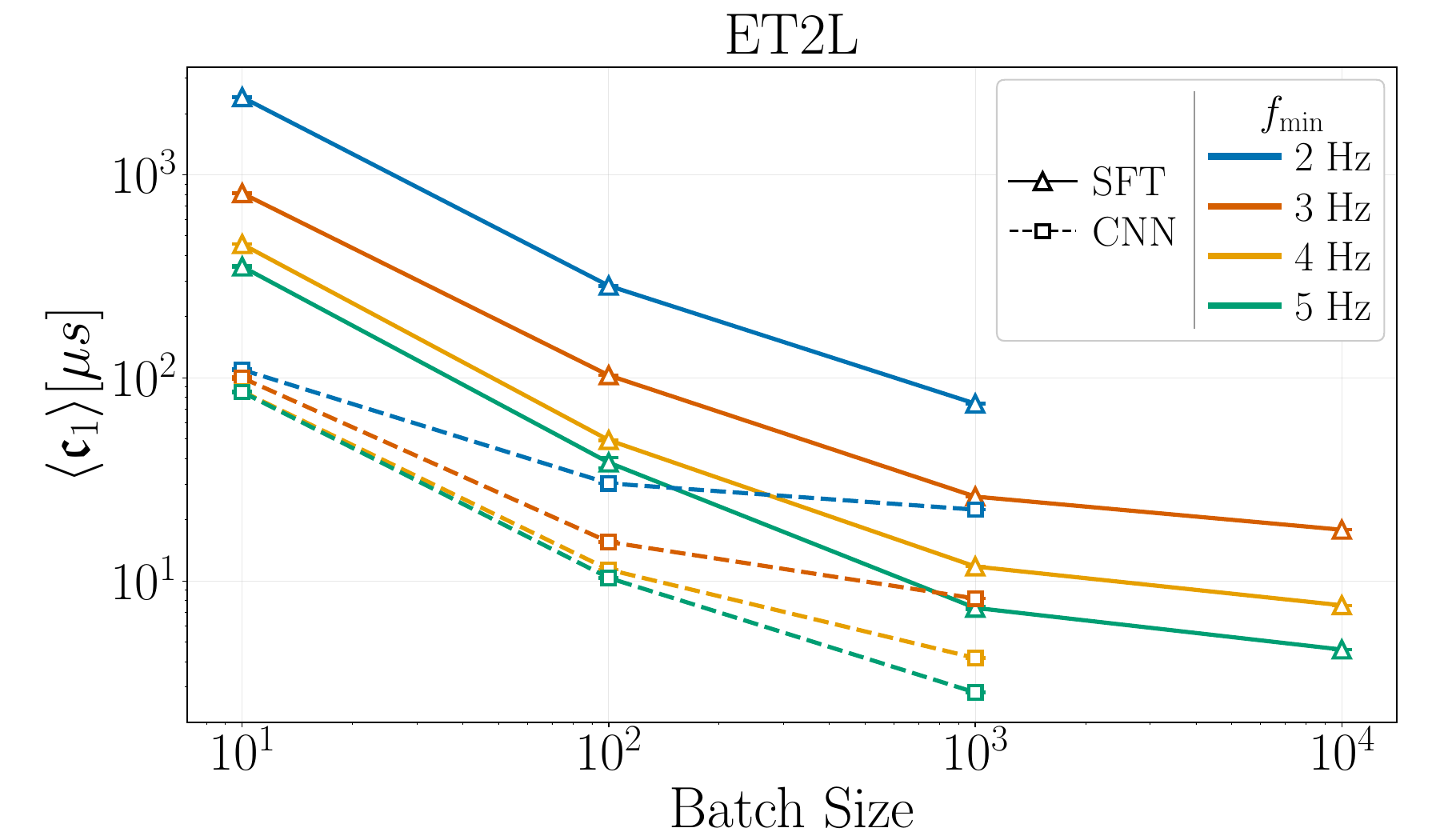}
    \caption{Average computational cost per template $\langle \mathfrak{c}_1 \rangle$ as a function of JAX's \texttt{vmap} batch size for the $\ETT$ (top) and $\ETL$ (bottom) detector configurations.
The CNN pipeline (dashed lines) is evaluated up to a batch size equal to the template-bank size at which its sensitivity saturates ($\mathcal{N} = 10^3$).
The SFT matched-filtering baseline (solid lines) reaches VRAM saturation at batch sizes of $\mathcal{N} = 10^4$ for $f_{\min} \geq 3$ Hz and $\mathcal{N} = 10^3$ for $f_{\min} = 2$ Hz.}
    \label{fig:throughput_scaling}
\end{figure}

Figure~\ref{fig:throughput_scaling} shows $\costOne$ as a function of detector configuration, minimum frequency, and batch size.
For our CNN pipeline, peak performance is achieved by evaluating the full template bank ($\mathcal{N}_{\mathrm{CNN}} = 10^3$) in a single batch, yielding $\costOne \approx (3-25)\, \mu\mathrm{s}$.
The SFT approach reaches batch sizes of $10^4$ for $f_{\min} \geq 3$ Hz and $10^3$ for $f_{\min} = 2$ Hz before saturating the GPU resources, with $\costOne \approx (3-100)\, \mu\mathrm{s}$.
All latency measurements and speedup factors reported below are obtained using these peak-throughput configurations.

\begin{figure}
    \centering
    \includegraphics[width=1\linewidth]{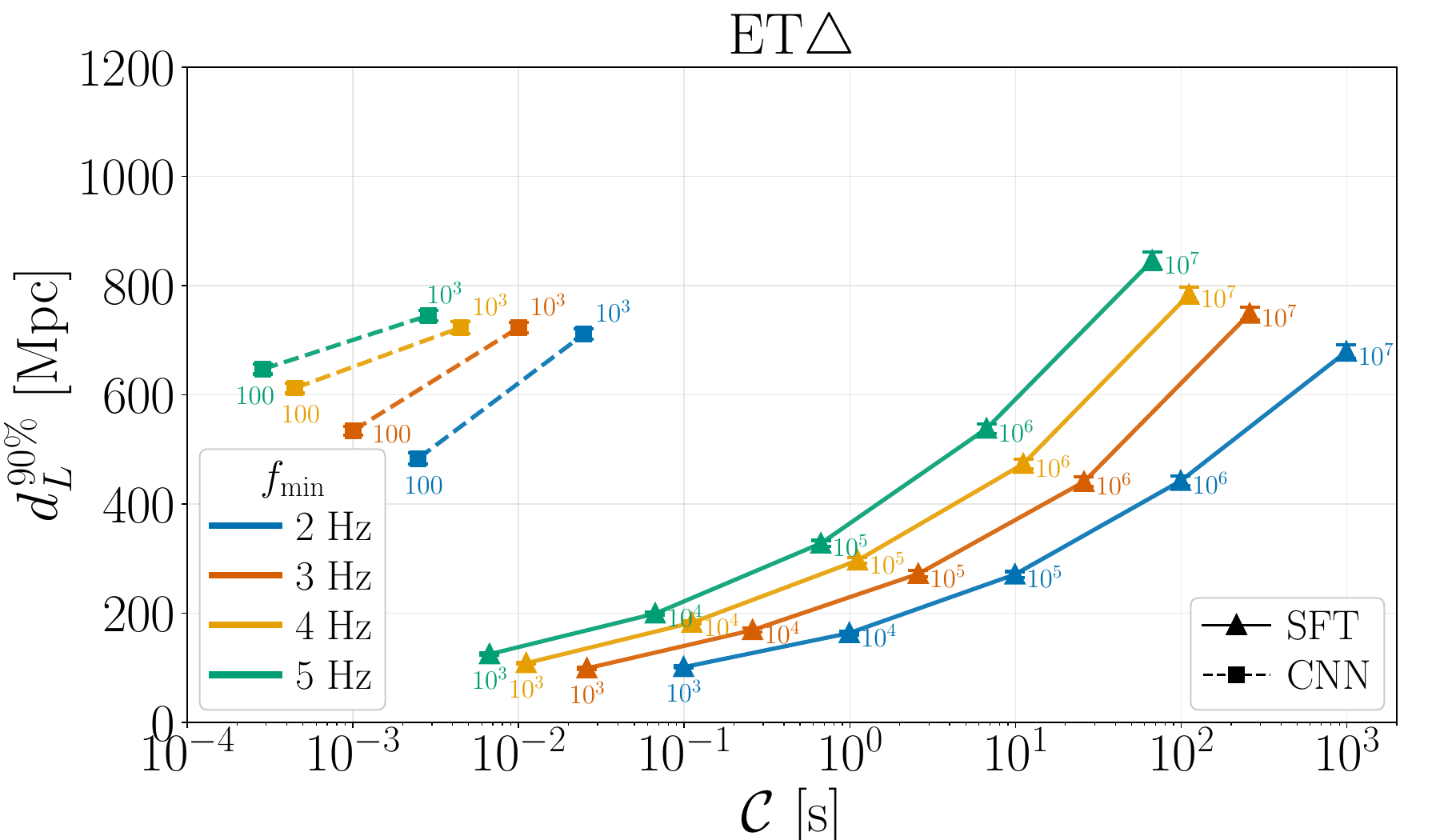}\\
    \vspace{0.4cm}
    \includegraphics[width=1\linewidth]{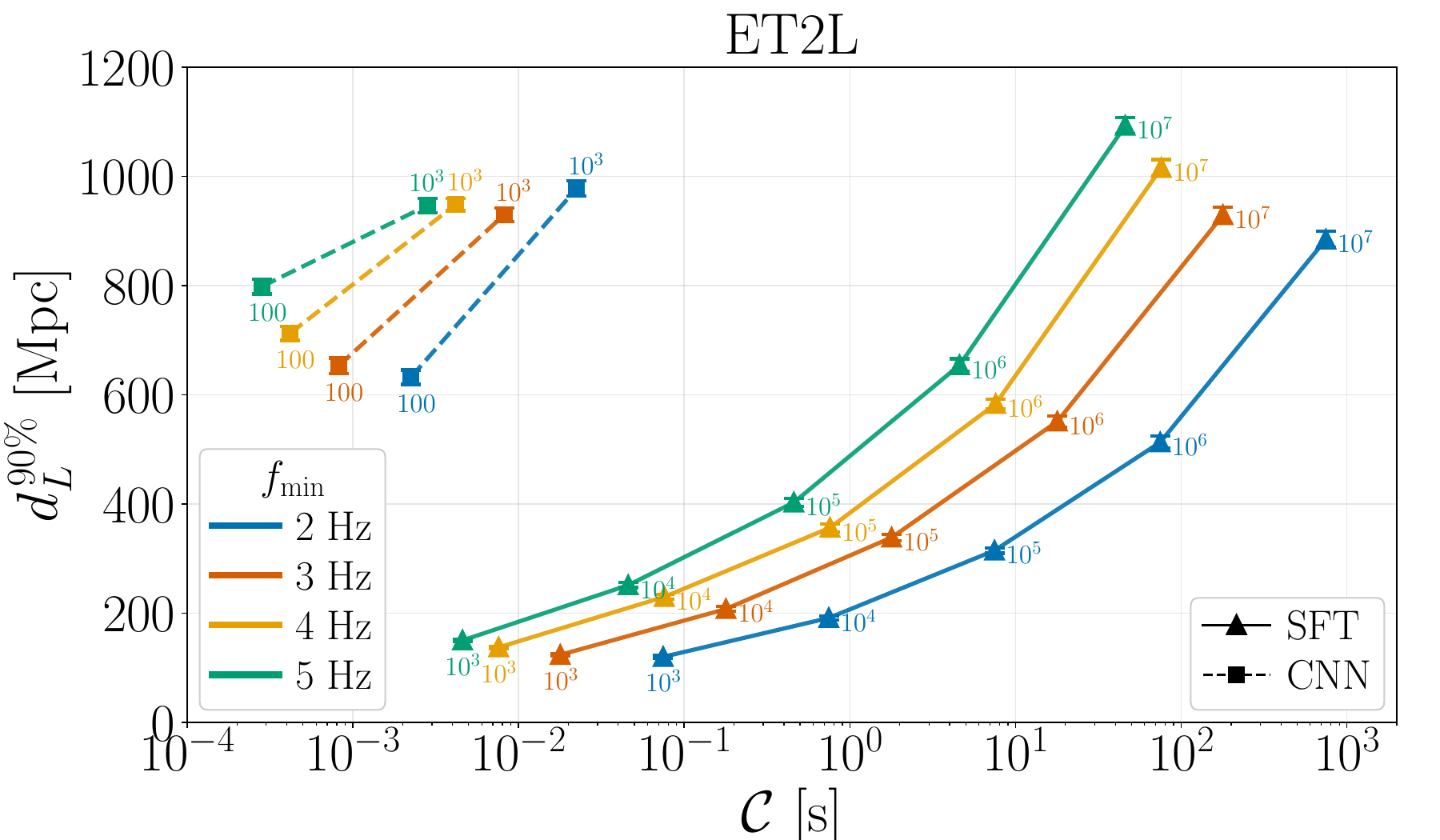}
    \caption{Detection horizon $d_L^{90\%}$ as a function of the evaluation latency $\cost$ for the ET$\triangle$ (top) and ET2L  (bottom) configurations. The performance of the proposed CNN pipeline (dashed lines, squares) is compared against the baseline SFT matched-filtering (solid lines, triangles) across four low-frequency regimes: $f_{\text{min}} = \{2,3,4,5\} \, \mathrm{Hz}$, while the maximum frequency is kept fixed to $f_\text{max} = 10 \ \text{Hz}$. Data points are annotated with their respective template bank size $\mathcal{N}$, extending up to $10^7$. The CNN reduces evaluation latency by over four orders of magnitude compared to the matched-filtering, achieving millisecond speeds, while reaching comparable sensitivity. 
    }
    \label{fig:efficiency_combined}
\end{figure}

Figure \ref{fig:efficiency_combined} shows the sensitivity horizon $d_L^{90\%}$ as a function 
of the latency $\cost$ for different template bank sizes. 
The maximum frequency is kept fixed to $f_\text{max} = 10 \ \text{Hz}$ in all configurations. 
Table \ref{tab:latency_results} shows evaluation 
times at the optimal number of templates, where
both methods achieve a comparable sensitivity.
Across all three frequency regimes and both ET configurations, 
our CNN using a bank of only $10^3$ templates, achieves detection capabilities comparable to
the SFT-based matched-filtering employing $10^7$ templates, 
while reducing evaluation times by over four orders of magnitude.

\begin{table}
    \begin{ruledtabular}
    \begin{tabular}{l c c c c}
        Config. & $f_{\min}$ & SFT & CNN & Gain \\
        \colrule
         \rule{0pt}{3ex}& 2 Hz & 12.4 min & 22.4 ms & $3.3 \times 10^4$ \\
        \hspace{0.2cm} 2L & 3 Hz & 3.0 min & 8.2 ms & $2.2 \times 10^4$ \\
        & 4 Hz & 1.3 min & 4.2 ms & $1.9 \times 10^4$ \\
         & 5 Hz & 0.8 min & 2.8 ms & $1.6 \times 10^4$ \\
        \colrule
         \rule{0pt}{3ex}& 2 Hz & 16.5 min & 24.8 ms & $4.0 \times 10^4$ \\
        \hspace{0.2cm} $\triangle$ & 3 Hz & 4.3 min & 10.0 ms & $2.6 \times 10^4$ \\
        & 4 Hz & 1.8 min & 4.5 ms & $2.4 \times 10^4$ \\
         & 5 Hz & 1.1 min & 2.9 ms & $2.3 \times 10^4$ \\
    \end{tabular}
    \end{ruledtabular}
    \caption{Evaluation latency comparison between the baseline SFT matched-filtering pipeline ($\mathcal{N}_{\text{SFT}}=10^7$) and CNN pipeline ($\mathcal{N}_{\text{CNN}}=10^3$). Execution times are calculated using the optimal batch size configuration for each method ($10^4$ for SFT matched-filtering at 3, 4, and 5 Hz, $10^3$ at 2 Hz due to VRAM saturation, and $10^3$ for CNN). The inter-run uncertainty across ten independent trials remains negligible ($\sim 10^{-2}\, \mathrm{ms}$ for CNN and $\sim 10^{-1} \,\mathrm{s}$ for matched-filtering).}
    \label{tab:latency_results}
\end{table}

\subsection{CNN dependency on $t_0$} \label{t0cnn}

The CNN is trained to distinguish between the signal and noise hypotheses.
A high detection probability alone, however, does not imply that it has learned the waveform morphology, as neural networks may instead exploit spurious features that distinguish the classes in the generated datasets~\cite{2016arXiv161103530Z,2020arXiv200407780G}.

\begin{figure}
    \centering
    \includegraphics[width=\columnwidth]{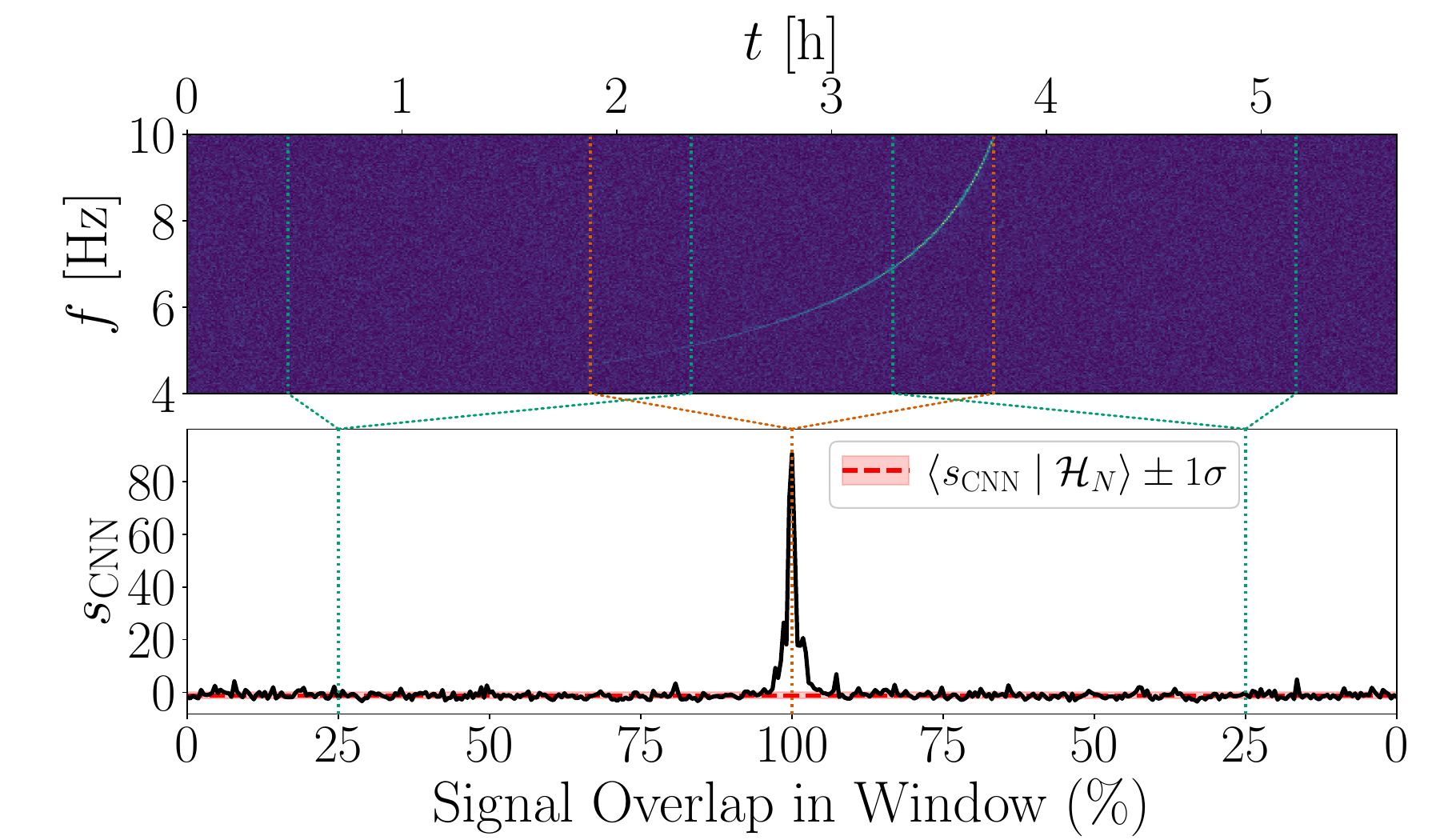}
    \caption{\emph{Upper}: Spectrogram of a BNS signal injected in $\approx 5.5$ hours of background stream. The luminosity distance is set to $d_L = 100 \, \mathrm{Mpc}$ for visualization purposes. 
    \emph{Lower}: Evaluation of the CNN detection statistic (logits) during a rolling window search on the data stream shown in the upper panel, with the system placed at $d_L = 400 \, \mathrm{Mpc}$. The horizontal axis tracks the percentage of the signal overlapping with the observation window, distinguishing between the entering and leaving phases.
    Dotted vertical lines mark the edges of three representative observation windows and connect them to their extent in the data stream above: two windows in green, containing 25\% of the signal as it enters and as it leaves, and one in orange, exactly aligned with the signal and therefore at 100\% overlap.
    The expected background distribution under the noise hypothesis $\mathcal{H}_N$ is shown in red, with the dashed line and shaded area indicating the mean and its $\pm 1\sigma$ interval, respectively.
   }
    \label{fig:rolling_window}
\end{figure}

To verify that the CNN responds to the GW signal itself, we evaluate its output on a data stream containing a BNS signal while varying the reference time $t_0$. In other words, we evaluate the CNN response to a BNS signal crossing the observation window. Figure~\ref{fig:rolling_window} shows the network output as the observation window slides across a continuous stream of background noise which contains one BNS signal injected with component masses $m_1 = m_2 = 1.4 \, \mathrm{M}_\odot$ at a luminosity distance of $d_L = 400 \, \mathrm{Mpc}$. As the signal enters the window, the signal overlap increases, and the detection statistic peaks only when the signal is fully contained within the observation window, and decreases rapidly as the signal becomes misaligned and leaves the window.
This behavior indicates that the network relies on the specific time-frequency evolution of the fully aligned waveform to maximize its detection statistic, rather than simply responding to localized excess energy within the observation window.

\subsection{Sensitivity dependence on $f_{\mathrm{min}}$}
The results in Fig.~\ref{fig:efficiency_combined} consistently show that for matched-filtering, for a fixed number of templates, the sensitivity distance increases with the minimum frequency $f_{\mathrm{min}}$.
In other words, a search with $f_{\mathrm{min}} = 5 \, \mathrm{Hz}$ is more sensitive than one with either \mbox{$f_{\mathrm{min}} = 2 \, \mathrm{Hz}$}, \mbox{$f_{\mathrm{min}} = 3 \, \mathrm{Hz}$}, or \mbox{$f_{\mathrm{min}} = 4 \, \mathrm{Hz}$}.
One might expect sensitivity to improve as the search extends to lower frequencies; below, we explain this apparent discrepancy using $\mathcal{N} = 10^7$ as an example.

\begin{figure}
     \centering
     \includegraphics[width=\columnwidth]{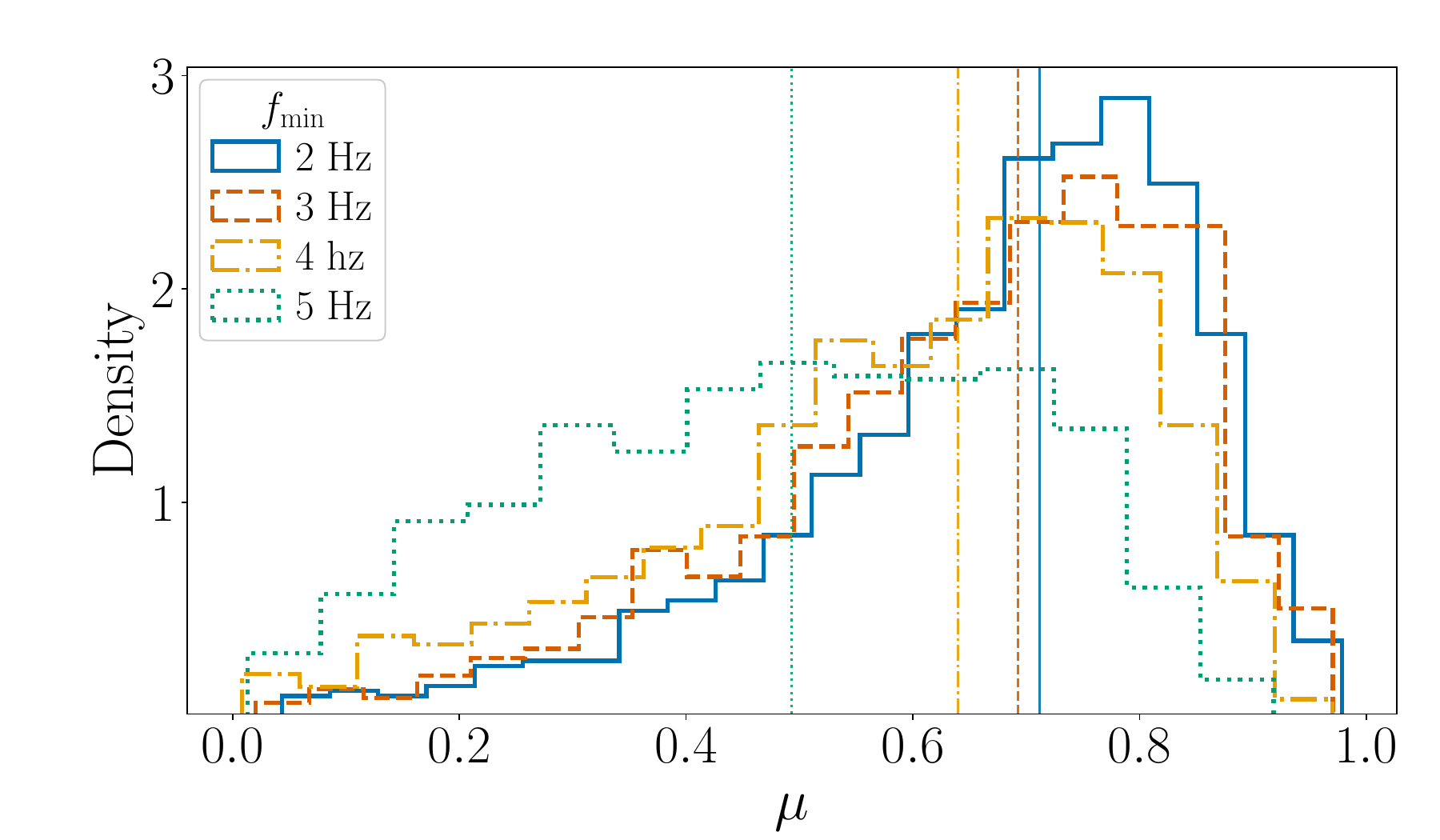}
     \caption{
Mismatch distributions for the $\ETT$ configuration, obtained from $10^3$ injections at each minimum frequency. Vertical lines mark the median values.
     }
     \label{fig:mismatch_freqs}
\end{figure}

Coherent detection statistics, such as those being approximated in this work, 
scale with the squared signal-to-noise ratio (SNR) of the signal~\cite{Owen:1995tm,Wette:2016raf,Jaranowski:1998qm}
\begin{equation}
    \langle s\rangle \sim \rho^2_{\mathrm{opt}} ( 1 - \langle\mu\rangle)
\end{equation}
where $\rho^2_{\mathrm{opt}} = \langle h, h\rangle$ for a given signal $h(t)$ and $\langle \mu \rangle$ is the
expected mismatch given a template bank,
defined as the relative loss of detection statistic $s(\theta)$~\cite{Allen:2021yuy, Wette:2016raf, Owen:1995tm}:
\begin{equation}
    \mu(\theta;\theta_{\mathrm{true}}) = 1-\frac{s(\theta)}{s(\theta_\text{true})} \,.
\end{equation}
As a function of the detector's PSD, and assuming an early-inspiral stage,
the optimal SNR is governed by~\cite{Maggiore:2007ulw}
\begin{equation}
    \rho_\text{opt}^2 \propto  \int_{f_\text{min}}^{10\,\mathrm{Hz}}\mathrm{d}f \, \frac{f^{-7/3}}{S_n(f)} \,.
\end{equation}
Considering the $\ETT$ configuration, lowering $f_{\mathrm{min}}$ from $5 \, \mathrm{Hz}$ to $2 \, \mathrm{Hz}$ yields an increase 
in $\rho_\text{opt}^2$ of $\sim 12.03\%$. For $\mathcal{N} = 10^7$, the median mismatch increases from $\sim 0.49$ to $\sim 0.71$, as shown in  Fig.~\ref{fig:mismatch_freqs}. 
Overall, the expected detection statistic decreases to $1.1203 \times (1 - 0.71) / (1 - 0.49) \approx 64\%$ of its original value.

Therefore, for the sensitivity curves~\cite{Danilishin:2023ETcurves} used in this study, the gain in $\rho_{\text{opt}}^2$ from longer signals is insufficient to offset the increased mismatch caused by the expansion of the parameter space.

\section{Conclusion}
The extended low-frequency sensitivity of 3G GW observatories significantly increases the duration of BNS signals. This breaks the short-signal approximation, requiring the inclusion of time-dependent amplitude and phase modulations and expanding the search space to include extrinsic parameters. As a result, standard matched-filtering approaches face a dimensionality problem: maintaining an acceptable mismatch requires a computationally intractable number of templates, which is incompatible with the latency requirements of early-warning applications.

To address this, we put forward an approach that integrates a template bank strategy with deep learning. Instead of replacing the entire detection pipeline with an end-to-end neural network, we leverage deep learning to marginalize the Bayes Factor over nuisance parameters and we use a chirpmass-only template bank to maximize it. The CNN is therefore employed exclusively to approximate the intractable marginalization over the extrinsic parameters by analyzing localized time-frequency crops of complex SFTs.

This strategy reduces the computational cost by over four orders of magnitude. Compared to a GPU-optimized SFT-based matched-filtering baseline that requires $10^7$ templates, our pipeline evaluates only $10^3$ templates while achieving comparable detection sensitivity. Consequently, total evaluation latency is reduced from minutes to milliseconds, providing a computationally viable framework for BNS early-warning in 3G networks.

Future developments will need to address the specific challenges of realistic 3G data streams, such as robustness against non-stationary instrumental noise and glitches. Furthermore, the extended duration of the signals implies a high probability of overlapping events \cite{Samajdar:2021egv,Baka:2025yqx,Papalini:2025exy}, requiring the network to be adapted to isolate individual tracks.
\begin{acknowledgments}

We thank Marco Bianchi, Justin Janquart, and Francesco Iacovelli for discussions.
M. G. is supported by the Fonds dédié aux actions de recherche concertée (ARC) under Grant 25-30-150 - GRAVIT-AI.
R.~T.~and D.~G.~are supported by
ERC Starting Grant No.~945155--GWmining, 
Cariplo Foundation Grant No.~2021-0555, 
Italian-French University (UIF/UFI) Grant No.~2025-C3-386,
MUR Grant ``Progetto Dipartimenti di Eccellenza 2023-2027'' (BiCoQ),
and the INFN TEONGRAV initiative.
D.G. is supported by MUR Young Researchers Grant No. SOE2024-0000125. 
Computational work was performed 
at CINECA with allocations through INFN and the University of Milano-Bicocca, 
and at NVIDIA with allocations through the Academic Grant program.

\end{acknowledgments}

\appendix
\section{Signal Model}
\label{app:signal_model}
Let us define a uniform temporal grid $t$ in the detector frame
\begin{equation}
    t_j = t_{\text{start}} + j \cdot \Delta t, \quad j = 0, \dots, N-1\,,
\end{equation}
where $\Delta t = 1 / f_{\text{s}}$ is the inverse of the sampling frequency, and $N$ is the number of samples.

Generating the detector response requires evaluating the waveform phase in an inertial frame shared by the source and the observer. We adopt the solar system barycenter (SSB) as the origin of this reference frame and map each detector timestamp $t_j$ to the corresponding SSB coordinate time $t_{\text{SSB}}$. This transformation accounts for the R{\o}mer delay arising from Earth's motion relative to the SSB. We neglect the Einstein and Shapiro delays, whose contributions are subdominant for the sensitivity studies considered here~\cite{Jaranowski:1998qm,Leaci:2015bka}.
The relation between the detector time $t$ and the SSB time $t_{\text{SSB}}$ is given by:
\begin{equation}
    t_{\text{SSB}}(t) = t + \frac{{r}_{\text{det}}(t)\,  {n}(\alpha, \delta)}{c},
\end{equation}
where the detector position vector $r_{\text{det}}$ with respect to the SSB at time $t$ is defined as the sum of the Earth's orbital motion and the detector's rotational motion relative to the Earth's center, assumed to be circular motions for simplicity~\cite{Jaranowski:1998qm}
\begin{equation}
  r_{\text{det}}(t) = r_{\text{orb}}(t) + r_{\text{rot}}(t),  
\end{equation}
and $n(\alpha, \delta)$ is the vector pointing to the source
\begin{equation}
    n = (\cos\alpha\cos\delta, \sin\alpha\cos\delta, \sin\delta).
\end{equation}

For our study, we assume the signal to be correctly centered within the observation window. The start time $t_{\text{start}}$ is determined by first identifying the end of the signal in the SSB frame and then projecting it to the detector frame.
Let $t_{\text{max}}^{\text{SSB}}$ be the time at which the signal reaches the maximum frequency $f_{\text{max}}$ in the SSB frame. We estimate the arrival time at the detector by correcting for the geometric delay:
\begin{equation}
    t_{\text{end}}^{\text{det}} \simeq t_{\text{max}}^{\text{SSB}} - \Delta t_{\text{R{\o}mer}}(t_{\text{max}}^{\text{SSB}})\,,
\end{equation}
where the R{\o}mer delay is approximated using the detector position at the merger time:
\begin{equation}
    \Delta t_{\text{R{\o}mer}} = \frac{r_{\text{det}}(t_{\text{max}}^{\text{SSB}}) \, n(\alpha, \delta)}{c}\,.
\end{equation}
Finally, the grid start time is set to capture the last  portion of the inspiral:
\begin{equation}
    t_{\text{start}} = t_{\text{end}}^{\text{det}} - N  \Delta t\,.
    \label{t_start}
\end{equation}

Once we compute the re-mapped timestamp vector $t_\text{SSB}(t)$, we can get the two polarization $h_+$ and $h_{\times}$ in the SSB frame as~\cite{Maggiore:2007ulw}:
\begin{align}
    h_+(t_\text{SSB}) &= A(t_\text{SSB}) \left(\frac{1+\cos^2\iota}{2}\right) \cos\phi(t_\text{SSB})\,,\\
    h_\times(t_\text{SSB}) &=A(t_\text{SSB})\cos\iota \sin\phi(t_\text{SSB})\,,
\end{align}
where $\iota$ is the inclination angle. The waveform's amplitude $A(t_\text{SSB})$ and phase 
$\phi(t_\text{SSB})$ depend on the binary's intrinsic parameters $\{m_1, m_2, \chi_{1L}, \chi_{2L}\}$
and a the luminosity distance $\dL$.

The signal measured at the detector is a combination of both polarizations, weighted by the response of the 
detector's antenna-pattern functions $F_{+, \times}(t;n,\psi)$, which depend both on the source's sky position and
polarization angle $\psi$
\begin{equation}
    \left[ \begin{array}{c} F_{+}(t;n, \psi) \\ F_{\times}(t; n,\psi)\end{array}\right] = 
    \sin\zeta
    \left[ \begin{array}{cc} \cos2\psi & \sin2\psi \\ -\sin2\psi & \cos2\psi\end{array}\right]
    \left[ \begin{array}{c}a(t;n) \\ b(t; n)\end{array}\right] \,,
\end{equation}
and the functions $a(t; n)$ and $b(t; n)$ represent the response of the detector to the two GW
polarizations~\cite{Jaranowski:1998qm}. The resulting time-domain strain is
\begin{equation}
    h(t) = F_+ (t)\  h_+\left[t_\text{SSB}(t)\right] + F_\times (t)\ h_\times\left[t_\text{SSB}(t)\right] \,.
    \label{eq:h_of_t}
\end{equation}

\section{Network architectures}
\label{appendix:CNNS}
The CNN architectures are summarized in Tables~\ref{tab:cnn_architecture_base} and \ref{tab:cnn_params}. The input length depends on the minimum frequency, while the number of input channels depends on the detector configuration: four for $\ETL$ and six for $\ETT$. All models were trained on an NVIDIA A100-SXM-64GB GPU node for $t_{\text{train}} \leq 24$ hours using the AdamW optimizer and a cosine learning-rate decay schedule with warmup. For $f_{\min} \in \{3, 4, 5\}$ Hz, the networks were trained on data generated on the fly for approximately $N_{\text{step}} \approx 10^6$ steps with a batch size of 128, corresponding to approximately $130 \times 10^6$ unique training samples. Validation was performed for 200 steps every $27\,000$ training steps, yielding approximately $10^6$ validation samples. For $f_{\min} = 2$ Hz, the models were trained on approximately $50 \times 10^6$ samples and evaluated on approximately $10^5$ validation samples.

\begin{table*}
    \centering
    \small
    \renewcommand{\arraystretch}{1.1} 
    \begin{tabular}{l l c c l}
        \hline\hline
        Block & Layer / Operation & Kernel $(K_H, K_W)$ & Stride $(S_H, S_W)$ & Output Shape $(C, H, W)$ \\
        \midrule
        
        Input & Input Tensor & - & - & $(2 N_{\mathrm{det}}, 51, W_0)$ \\
        \midrule
        
        \multirow{2}{*}{Conv 1}
          & Conv2D & $(2, K_{W1})$ & $(1, 2)$ & $(32, 50, W_1)$ \\
          & ReLU & - & - & $(32, 50, W_1)$ \\
        \midrule
        
        \multirow{3}{*}{Conv 2} 
          & Conv2D & $(5, K_{W2})$ & $(3, S_{W2})$ & $(64, 16, W_2)$ \\
          & BatchNorm & - & - & $(64, 16, W_2)$ \\
          & ReLU & - & - & $(64, 16, W_2)$ \\
        \midrule
        
        \multirow{3}{*}{Conv 3}
          & Conv2D & $(5, 5)$ & $(3, S_{W3})$ & $(128, 4, W_3)$ \\
          & BatchNorm & - & - & $(128, 4, W_3)$ \\
          & ReLU & - & - & $(128, 4, W_3)$ \\
        \midrule
        
        \multirow{3}{*}{Conv 4} 
          & Conv2D & $(2, K_{W4})$ & $(2, K_{W4})$ & $(256, 2, W_4)$ \\
          & BatchNorm & - & - & $(256, 2, W_4)$ \\
          & ReLU & - & - & $(256, 2, W_4)$ \\
        \midrule
        
        Flatten & Ravel & - & - & $(N_{\text{flat}})$ \\
        \midrule
        
        \multirow{5}{*}{Classifier} 
          & Dropout ($p=0.5$) & - & - & $(N_{\text{flat}})$ \\
          & Linear (Dense) & - & - & $(128)$ \\
          & ReLU & - & - & $(128)$ \\
          & Dropout ($p=0.5$) & - & - & $(128)$ \\
          & Linear (Output) & - & - & $(1)$ \\
        \hline\hline
    \end{tabular}
        \caption{Adopted CNN architecture. The height dimension ($H$) processing is constant, while the width dimension ($W$, time axis) parameters depend on $f_{\text{min}}$ and are detailed in Table \ref{tab:cnn_params}.}
    \label{tab:cnn_architecture_base}

\end{table*}

\begin{table*} %
    \centering
    \small 
    \renewcommand{\arraystretch}{1.2} 
    \begin{tabular}{c | c | c c | c c | c c | c c | c}
        \hline \hline
        \multirow{2}{*}{$f_{\text{min}}$} & Input & \multicolumn{2}{c|}{Conv 1} & \multicolumn{2}{c|}{Conv 2} & \multicolumn{2}{c|}{Conv 3} & \multicolumn{2}{c|}{Conv 4} & Flatten \\
         & $W_0$ & $K_{W1}$ & $W_1$ & $(K_{W2}, S_{W2})$ & $W_2$ & $S_{W3}$ & $W_3$ & $K_{W4}$ & $W_4$ & $N_{\text{flat}}$ \\
        \midrule
        {2 Hz} & 660 & 3 & 329 & $(5, 4)$ & 82 & 4 & 20 & 5 & 4 & 2048 \\
        {3 Hz} & 218 & 2 & 109 & $(5, 3)$ & 35 & 4 &  8 & 2 & 4 & 2048 \\
        {4 Hz} &  96 & 2 &  48 & $(6, 3)$ & 14 & 3 &  4 & 2 & 2 & 1024 \\
        {5 Hz} &  49 & 2 &  24 & $(5, 3)$ &  7 & 2 &  2 & 2 & 1 &  512 \\
        \hline \hline
    \end{tabular}
        \caption{Time-axis ($W$) hyperparameters and layer output widths corresponding to different $f_{\text{min}}$ cutoffs .}
    \label{tab:cnn_params}

\end{table*}

\section{SFT-based Matched Filtering}
\label{sft}

Searches using the dominant mode construct a detection statistic using
the log-likelihood~\cite{Harry:2010fr,Capano:2013raa,Usman:2015kfa,Capano:2016dsf}
\begin{equation}
    \mathcal{L} = \langle d, h\rangle - \frac{1}{2}\langle h, h\rangle \,,
\end{equation}
where  we omitted signal-independent terms and used the standard scalar product~\cite{Finn:1992wt}
\begin{equation}
    \langle a, b \rangle = 4 \mathrm{Re}\int_{0}^{\infty} \mathrm{d}f
    \frac{\tilde{a}^{*}(f) \tilde{b}(f)}{S_{\mathrm{n}}(f)} \,.
\end{equation}
For computational efficiency, search strategies typically remove the dependency on the initial phase
and distance by using a maximum likelihood estimate that can be computed in closed form. 
Let us write the dependency of the signal model [Eq.~\eqref{eq:h_of_t}] on $\dL$ and 
$\phi_0$ as 
\begin{equation}
    h(t) = \frac{ h_{\mathrm{c}}(t)\cos\phi_0 + h_{\mathrm{s}}(t)\sin\phi_0}{\dL} \,,
\end{equation}
where
\begin{align}
    h_{\mathrm{c}}(t) &= \Ap(t) \cos\phi(t) + \Ac(t)\sin\phi(t)  \label{eq:h_c} \,,\\
    h_{\mathrm{s}}(t) &= \Ac(t) \cos\phi(t) - \Ap(t) \sin\phi(t) \,,
    \label{eq:h_s}
\end{align}
and
\begin{align}
    \Ap(t) &= A(t)\frac{1 + \cos^2 \iota}{2}F_+(t)\,, \\
    \Ac(t) &=A(t)\cos\iota F_\times(t) \,.
\end{align}
The log-likelihood $\mathcal{L}$ is
\begin{equation}
    \mathcal{L}
    = \frac{ \langle d, h_{\mathrm{c}}\rangle \cos\phi_0 + \langle d, h_{\mathrm{s}}\rangle \sin\phi_0 }{\dL} - \frac{\langle h_c, h_c\rangle}{2\dL^2} \,. 
\end{equation}
where $\dL^2\langle h, h\rangle \simeq \langle h_{\mathrm{c}}, h_{\mathrm{c}}\rangle  \simeq \langle h_{\mathrm{s}}, h_{\mathrm{s}}\rangle$. 
Maximization with respect to $\phi_0$ yields
\begin{equation}
    \max_{\phi_0} \mathcal{L} = 
    \frac{\sqrt{\langle d, h_{\mathrm{c}} \rangle^2 + \langle d, h_{\mathrm{s}} \rangle^2 }}{\dL} 
    - \frac{\langle h_c, h_c\rangle}{2\dL^2} \,,
\end{equation}
and further maximizing with respect to $\dL$ returns the expected result
\begin{equation}
     \mathcal{X}\equiv \max_{\phi_0, \dL} \mathcal{L} = \frac{\langle d, h_{\mathrm{c}} \rangle^2 + \langle d, h_{\mathrm{s}} \rangle^2 }{2 \langle h_c, h_c\rangle}\,.
     \label{eq:mf_det}
\end{equation}
Finally, the detection statistic for a search with $\mathcal{N}$ templates is given by:
\begin{equation}
  s_{\mathrm{SFT}} = \max_{j \in \{1, \dots, \mathcal{N}\}}  \mathcal{X}(\theta_j)  \,.
\end{equation}

We restrict the  analysis to  signals evolving between $f_{\text{min}} \in \{2, 3, 4, 5\}$
Hz and $f_{\text{max}} = 10\,\mathrm{Hz}$. To strictly satisfy the Nyquist-Shannon sampling theorem ($f_s > 2f_{\text{max}}$) and avoid aliasing artifacts, the sampling frequency is set to $f_s = 40$ Hz for the 3 Hz, 4 Hz, and 5 Hz regimes, and reduced to $f_s = 25$ Hz for the 2 Hz scenario, to mitigate the increased computational load associated with longer waveforms.

These scalar products can be efficiently computed using the
SFT basis as presented in Ref.~\cite{Tenorio:2025gci}.
Concretely, Eqs.~\eqref{eq:h_c} and~\eqref{eq:h_s} can be recast in terms of a projector
\begin{align}
    h_{\mathrm{c/s}}(t) &=\Lambda_{\mathrm{c/s} ,t} e^{i \phi(t)} \,,\\
    \Lambda_{\mathrm{c} ,t} &= \Ap(t) \mathrm{Re} + \Ac(t)\mathrm{Im}\,,\\ 
    \Lambda_{\mathrm{s} ,t} &= \Ac(t) \mathrm{Re} - \Ap(t)\mathrm{Im} \,.
\end{align}

The scalar product is therefore formulated as
\begin{align}
    ([d, h_{\mathrm{c/s}}]) &= 2 \sum_{\alpha=0}^{N_{\text{SFT}}-1} 
    \Lambda_{\mathrm{c/s},t_{\alpha}} e^{i \phi_{\alpha}} C(f_{\alpha}, \dot{f}_{\alpha};\tilde{d}_{\alpha}) \,, \label{eq:sft_scalar}
    \end{align}
    \begin{align}
    C(f_\alpha, \dot{f}_\alpha; \tilde{d}_\alpha) &=     \Delta f\sum_{k=k_{\min}}^{k_{\max}} \frac{\tilde{d}_\alpha^*[k]}{S_{\mathrm{n}}(f_k)} 
    \mathfrak{Fre}(f_\alpha - f_k, \dot{f}_\alpha),\\ 
    \mathfrak{Fre}(f_0, f_1) &= 
    \Delta t \sum_{j=0}^{n_{\text{SFT}}-1} e^{i(2\pi f_0 \tau_j + \pi f_1 \tau_j^2)} \,,
\end{align}
where $\alpha$ labels the SFT corresponding to timestamp \mbox{$t_{\alpha} = t_0 + \alpha \Tsft$}
and $\tilde{d}_{\alpha}(f)$ is the SFT of the data following the convention of Eq.~\eqref{eq:h_SFT}.
The frequency index $k$ runs over $\Delta k = k_{\mathrm{max}} - k_{\mathrm{min}} = 51$ 
frequency bins around the instantaneous frequency of the signal $f_{\alpha}$.

The matched-filter for a given template $\theta$ corresponds to evaluating the projection of data $d(t)$ onto the corresponding waveform $h_{\mathrm{c,s}}(t)$ using Eq.~\eqref{eq:sft_scalar}, and normalizing by the exact bin-by-bin whitened template norm:
\begin{equation}
\begin{split}
    &([ h_{\mathrm{c}}, h_{\mathrm{c}}]) \approx\\ &\sum_{\alpha} \left[ \Ap^2(t_{\alpha})+ \Ac^2(t_{\alpha}) \right]
    \left( \Delta f \sum_{k=k_{\min}}^{k_{\max}} \frac{|\mathfrak{Fre}(f_\alpha - f_k, \dot{f}_\alpha)|^2}{S_{\mathrm{n}}(f_k)} \right)
\end{split}
\end{equation}
where we neglected rapidly oscillating terms between the + and $\times$ polarizations. 

\begin{table}[t]
    \centering
    \renewcommand{\arraystretch}{1.4} %
    \begin{tabular}{lcc}
        \hline \hline
        Physical Quantity & Symbol & Prior Distribution \\
        \hline
        Component masses & $m_{1,2}$ & $\mathcal{U}(1.0 \, M_\odot, 3.0 \, M_\odot)$ \\
        Aligned spins & $\chi_{1L,2L}$ & $\mathcal{U}(-0.05, 0.05)$ \\
        Right ascension & $\alpha$ & $\mathcal{U}(0, 2\pi)$ \\
        Declination & $\sin\delta$ & $\mathcal{U}(-1, 1)$ \\
        Polarization angle & $\psi$ & $\mathcal{U}(-\pi/2, \pi/2)$ \\
        Inclination & $\cos\iota$ & $\mathcal{U}(-1, 1)$ \\
        \hline \hline
    \end{tabular}
    \caption{Prior distributions defining the 8-dimensional parameter space $\theta$ used for the template bank generation.} 
    \label{tab:template_priors}
\end{table}

\begin{table}
    \begin{ruledtabular}
    \begin{tabular}{c c r c}
        ${f_{\min}}$ {[Hz]} & ${f_s}$ {[Hz]} & \multicolumn{1}{c}{${N}$} & ${T_{\text{sig}}}$ {[s]} \\
        \colrule
        \rule{0pt}{3ex}2 & 25 & 511\,500 & 20\,460 \\
        3 & 40 & 270\,320 & 6\,758 \\
        4 & 40 & 119\,040  & 2\,976 \\
        5 & 40 & 60\,760  & 1\,519 \\
    \end{tabular}
    \end{ruledtabular}
    \caption{Total number of samples $N$ and resulting signal duration $T_{\text{sig}}$ for each low-frequency cutoff $f_{\min}$ evaluated in this work. The durations are derived using the specific sampling frequencies $f_s$ configured for each regime.}
    \label{tab:signal_durations}
\end{table}

This results  extends seamlessly to a network of $N_{\mathrm{det}}$ detectors by reinterpreting
the scalar product
$\langle \cdot, \cdot\rangle$ in Eq.~\eqref{eq:mf_det} as a multi-detector scalar product.
For the case of uncorrelated detectors discussed in this work, the multi-detector scalar product
is the sum of single-detector scalar products~\cite{Cutler:2005hc}, which is equivalent to
re-interpreting $\alpha$ as an index running along {all} SFTs from all detectors ($X=1, \dots N_{\mathrm{det}}$):
\begin{equation}
    ([ d , h ]) = \sum_{X=1}^{N_{\mathrm{det}}} ([ d_{X}, h_{X} ]) \,.
\end{equation}

We construct a random template bank using the parameter distributions summarized in Table \ref{tab:template_priors}.
The waveform components $A_\alpha, f_\alpha, \dot{f}_\alpha$, and $\phi_\alpha$ are evaluated
using \texttt{iphenot}~\cite{Tenorio:2025gci}.

We fix the segment duration strictly to $T_{\text{SFT}} = 31$ s. This configuration optimally balances the computational efficiency of the SFT network with the strict phase requirements of the matched-filtering baseline.

Given the sampling frequency $f_s = 40$ Hz and the chosen segment duration $T_{\text{SFT}} = 31$ s, we can precisely define the spectral dimensions of our data for the $f_\text{min} = 3$ Hz, 4 Hz and 5 Hz regimes. Because the GW strain is a real-valued time series, computing the real-valued Fast Fourier Transform yields a physical, single-sided spectrum. The total number of discrete frequency bins is thus given by:
\begin{equation}
   N_{\text{bins}} = \frac{f_s \, T_{\text{SFT}}}{2} + 1 = \frac{40 \times 31}{2} + 1 = 621.
   \label{eq:n_bins}
\end{equation}
These 621 bins uniformly cover the physical frequency range from 0 Hz up to the Nyquist frequency 
$f_{\text{Nyq}} = 20$ Hz, resulting in a fine frequency resolution of $\Delta f = 1/T_{\text{SFT}} \approx 0.032$ Hz.
For the $f_\text{min} = 2$ Hz scenario one has:
\begin{equation}
   N_{\text{bins}} = \frac{f_s \, T_{\text{SFT}}}{2} + 1 = \frac{25 \times 31}{2} + 1 = 388.
   \label{eq:n_bins_2hz}
\end{equation}

\section{Using JAX on signal with different durations}
\label{appendix:signal_lenght}
The simulation pipeline is implemented using JAX~\cite{jax2018github} to leverage hardware acceleration. 
JAX requires fixed array shapes at execution to avoid re-compilations.
Consequently, we must enforce a uniform, fixed temporal length for all generated waveforms, 
regardless of the physical duration dictated by the specific binary parameters. 

To determine this fixed grid size, we use as a reference the shortest possible signal within our prior volume, which corresponds to the system with the highest chirp mass ($m_1 = m_2 = 3 \, M_\odot$). Setting the observation window to match this exact duration ensures that the fastest-evolving signal in the parameter space perfectly fits the array without introducing undefined values (NaNs) during generation.

For binaries with lower chirp masses, whose frequency evolution from $f_\text{min}$ to $f_\text{max}$ takes longer than this fixed window, we align the signals at the maximum frequency $f_{\text{max}}$ and extract only the final portion of the inspiral, as seen in Eq.~(\ref{t_start}). For the frequency regimes analyzed in this paper, the fixed grid sizes $N$ and their corresponding total observation durations $T_{\text{sig}}$ are summarized in Table~\ref{tab:signal_durations}.

\bibliographystyle{apsrev4-2} 
\bibliography{3gwarning}

\end{document}